\documentclass[twocolumn]{aastex631}

\usepackage{amsthm,amsmath,amssymb,mathrsfs,mathtools}
\usepackage{float}
\usepackage{enumitem}
\usepackage{multirow}
\usepackage{comment}
\usepackage{soul}

\shorttitle{CLU Planetary Nebulae}
\shortauthors{Du et al.}
\graphicspath{{./}{figures/}}

\begin{document}

\title{Prospects for Systematic Planetary Nebulae Detection with the Census of the Local Universe Narrowband Survey}

\correspondingauthor{Rong Du}
\email{durong@alumni.pku.edu.cn}

\author[0009-0006-6543-6333]{Rong Du}
\affiliation{Cahill Center for Astronomy and Astrophysics, California Institute of Technology, Pasadena, CA 91125, USA}
\affiliation{IPAC, California Institute of Technology, 1200 E. California Boulevard, Pasadena, CA 91125, USA}
\affiliation{Kavli Institute for Astronomy and Astrophysics, Peking University, Beijing 100871, China}

\author[0000-0002-6877-7655]{David O. Cook}
\affiliation{IPAC, California Institute of Technology, 1200 E. California Boulevard, Pasadena, CA 91125, USA}\email{dcook@ipac.caltech.edu}

\author[0000-0003-2071-2956]{Soumyadeep Bhattacharjee}
\affiliation{Cahill Center for Astronomy and Astrophysics, California Institute of Technology, Pasadena, CA 91125, USA}

\author[0000-0001-5390-8563]{Shrinivas R. Kulkarni}
\affiliation{Cahill Center for Astronomy and Astrophysics, California Institute of Technology, Pasadena, CA 91125, USA}
\affiliation{Department of Astronomy, Cornell University, Ithaca, NY 14853, USA}

\author[0000-0002-4223-103X]{Christoffer Fremling}
\affiliation{Cahill Center for Astronomy and Astrophysics, California Institute of Technology, Pasadena, CA 91125, USA}

\author[0000-0001-6295-2881]{David L. Kaplan}
\affiliation{Department of Physics, University of Wisconsin-Milwaukee, PO Box 413, Milwaukee, WI 53201, USA}

\author[0000-0002-5619-4938]{Mansi M. Kasliwal}
\affiliation{Cahill Center for Astronomy and Astrophysics, California Institute of Technology, Pasadena, CA 91125, USA}

\author[0000-0003-2451-5482]{Russ R. Laher}
\affiliation{IPAC, California Institute of Technology, 1200 E. California Boulevard, Pasadena, CA 91125, USA}

\author[0000-0002-8532-9395]{Frank J. Masci}
\affiliation{IPAC, California Institute of Technology, 1200 E. California Boulevard, Pasadena, CA 91125, USA}

\author[0000-0003-4401-0430]{David L. Shupe}
\affiliation{IPAC, California Institute of Technology, 1200 E. California Boulevard, Pasadena, CA 91125, USA}

\author[0000-0002-0331-6727]{Chaoran Zhang}
\affiliation{Center for Gravitation, Cosmology, and Astrophysics, Department of Physics, University of Wisconsin-Milwaukee, PO Box 413, Milwaukee, WI 53201, USA}

\begin{abstract}
We investigate the efficacy of a systematic planetary nebula (PN) search in the Census of the Local Universe (CLU) narrowband (H$\alpha$) survey that covers a considerably larger sky region of above declination $-20^\circ$ than most previous surveys. Using PNe observed by the Isaac Newton Telescope Photometric H$\alpha$ Survey (IPHAS) as validation, we are able to visually recover 432 out of 441 cataloged PNe (98\%) within the CLU dataset, with 5 sources having unusable CLU images and 4 missed due to limitations of imaging quality. Moreover, the reference PNe are conventionally divided into three PN classes in decreasing order of identification confidence given their spectra and morphologies. We record consistently high recovery rate across all classes: 95\% of True, 71\% of Likely, and 81\% of Possible sources are readily recovered. To further demonstrate the ability of CLU to find new PNe, we undertake a preliminary search of compact PNe within a sub-region of the validation catalog, mainly utilizing the significance of narrow-band colors ($\Sigma$) as a metric for identification. In a $200\,\rm deg^2$ region, we search the CLU source catalog and find 31 PN candidates after automated and visual scrutiny, of which 12 are new sources not appearing in previous studies. As a demonstration of our ongoing follow-up campaign, we present medium-resolution optical spectra of six candidates and notice that four of them show emission signatures characteristic of confirmed PNe. As we refine our selection methods, CLU promises to provide a systematic catalog of PNe spanning $2/3$ of the sky.
\end{abstract}

\keywords{Planetary nebulae (1249), H II regions (694), Surveys (1671), H alpha photometry (691)}

\section{Introduction}
Planetary nebulae (PNe) represent the late phase in the life cycle of low- to intermediate-mass stars \citep[1 to 8 $M_\odot$; e.g.,][]{kwok2005}. They offer valuable insights into the short phase of stellar evolution, and the composition and related dynamics of the interstellar medium. The relatively brief evolutionary stage, spanning approximately 25,000 years \citep[e.g.,][]{jacob2013,badenes2015,le-du2022}, encapsulates the transition from the asymptotic giant branch to white dwarfs \citep{paczynski1971a,paczynski1971b}. The hot white dwarf ionizes the surroundings and aids in the expansion of the ejected stellar shell through its winds \citep[e.g.,][]{kwok1978,frank1990}. The ionized shell then forms a PN with robust nebular emission lines, whereas the white dwarf forms the central star of the PN (CSPN) with a temperature exceeding $\sim 10^4$ to even $10^5 \, \rm K$ \citep[e.g.,][]{frew2010}. The visibility of these emission lines over considerable distances allows for the determination of PNe sizes and expansion velocities, thereby facilitating more precise exploration of the time-scales associated with the ejection \citep[e.g.,][]{iben1995}. The elemental composition of the ejecta provide direct insights into stellar evolution and chemical enrichment in galaxies \citep[e.g.,][]{dopita1997,maciel2003}. On a cosmic scale, the bright-end exponential cut-off of the luminosity function of PNe serves as a standard candle for calibrating cosmological distances \citep[e.g.,][]{ford1978,ciardullo1989,jacoby1989,jacoby1997,ciardullo2002,scheuermann2022}.

The discovery of the first known PN in 1764 (``Dumbbell'' Nebula, Messier\,27) by Charles Messier marked the beginning of a quest to understand these enigmatic objects. Subsequent decades saw the compilation of the first catalogs of PNe by \citet{abell1966} and also \citet{perek1967} whose sources were robustly confirmed with spectroscopic studies. Recent discoveries, primarily propelled by extensive optical surveys focusing on the H$\alpha$ emission line, have substantially augmented the list of candidates as well as confirmed PNe. Examples include the Macquarie-AAO-Strasbourg H$\alpha$ (MASH) Survey Catalog \citep{parker2006,miszalski2008} derived from the AAO/UKST SuperCOSMOS H$\alpha$ survey \citep[SHS;][]{parker2005}, and the Isaac Newton Telescope (INT) Photometric H$\alpha$ Survey \citep[IPHAS;][]{drew2005}. Further contributions to the PN population come from surveys in the Galactic bulge region, such as those conducted by \citet{boumis2006}, \citet{gorny2006}, the Deep Sky Hunters consortium \citep[e.g.,][]{jacoby2010,kronberger2012,kronberger2014,kronberger2016}, and the amateur community \citep[e.g.,][]{acker2012}. These diverse discoveries have been collated into the Hong Kong/AAO/Strasbourg/H$\alpha$ (HASH) PN database \citep{parker2014,parker2016,bojicic2017}. 

Despite recent advancements that have amply increased the known Galactic PN population, a significant gap of potentially up to an order of magnitude remains when comparing the number of known PNe ($\sim 3900$) to the theoretically predicted number \citep[$\sim 6,600$ to 60,000; e.g.,][]{frew2005,moe2006,jacoby2010,frew2017}. This discrepancy may introduce bias in analysis and hinders attempts to characterize the global PN population accurately. Therefore, the discovery of new PNe is necessary to enhance the census of this essential celestial object and its luminosity function.

To efficiently expand the catalog of known PNe \citep[e.g., see][]{kwitter2022,parker2022}, we leverage imaging data from the Census of the Local Universe \citep[CLU;][]{cook2019} narrowband survey which was conducted as part of the intermediate Palomar Transient Factory \citep[iPTF;][]{law2009,rau2009} program. While the primary science case of CLU endeavors to survey the local Universe for galaxies with distances out to 200 Mpc ($z \sim 0.05$) across $2/3$ of the sky ($26,470 \, \rm deg^2$; see \citealt{cook2019} for details), the survey holds the potential in unveiling all types of galactic H$\alpha$ emission-line objects including PNe. With its large sky coverage, the survey is ideal for exploring sky regions that lie outside the footprint of previous surveys like SHS \citep{parker2005} and IPHAS \citep{drew2005} to contribute a homogeneous PNe sample spanning $2/3$ of the entire sky.

As a pilot study, we investigate the prospects of discovering new PNe with CLU in this paper. The rest of the paper is organized as follows. We introduce the CLU survey in Section~\ref{sec:clu}. In Section~\ref{sec:rediscov}, we test the efficacy of detecting PNe in CLU-H$\alpha$ images by constructing a validation catalog using PNe detectable in IPHAS and quantifying its recovery rate. To further demonstrate the ability of the CLU database, we conduct a preliminary search within the sky field of the validation catalog to find more PNe as described in Section~\ref{sec:preliminary}. We discuss the caveats of our PNe search and the future prospective of CLU in Section~\ref{sec:discussion}. Finally, we present our conclusions in Section~\ref{sec:conclusion}.

\section{The CLU H$\alpha$ Survey} \label{sec:clu}
As detailed in \citet{cook2019}, the CLU survey represents a comprehensive endeavor to catalog and characterize H$\alpha$ emitting galaxies across the sky above $-20^\circ$ declination. The survey is conducted on the 1.2\,m Oschin Telescope at Palomar Observatory, employing four adjacent narrowband H$\alpha$ filters (hereafter, H$\alpha$1--H$\alpha$4) with their central wavelengths at 6563.7\,\AA, 6642.7\,\AA, 6717.6\,\AA, and 6808.6\,\AA, respectively. The CLU survey is capable of detecting H$\alpha$ emission lines down to $1.6 \times 10^{-14}\rm \, erg~s^{-1}~cm^{-2}$ ($143.95\, \rm R$ with $2.5''$ photometric aperture radius) at 90\% completeness (\citealt{cook2019}, 2024; submitted). It has been successfully used for finding galaxies and has discovered several interesting objects like blue compact dwarfs, green peas \citep[e.g.,][]{lintott2008}, and Seyfert galaxies \citep{cook2019}.

The scope of the survey, however, need not be limited to its goal on galaxies. In fact, the survey design is apt to search for many galactic H$\alpha$ emission-line objects, with our work here focusing on PNe. Interestingly, alongside galaxies, a known PN\,\citep[H\,4-1;][]{haro1951} has also been recovered in one of the CLU preliminary fields \citep{cook2017,cook2019}. The magnitude, $m$, of the object in the rest-frame H$\alpha$ filter (``H$\alpha$1'' or ``On'' filter) compared to the adjacent filter (``H$\alpha$2'' or ``Off'' filter) is $m_\mathrm{H\alpha1} - m_\mathrm{H\alpha2} = -3.05$, which implies a very significant detection of this object in the CLU survey \citep{cook2019}. We note that this PN is situated at a high Galactic latitude of $b=88.14757^\circ$ and $l = 49.3065^\circ$, which is outside the footprint of the previous major surveys, thus supporting the effectiveness of CLU in exploring these areas in the sky. Its measured H$\alpha$-line flux and equivalent width from the CLU data are $1.21 \times 10^{-12} \rm \, erg s^{-1}~cm^{-2}$ and 1200\,\AA, respectively.

\begin{deluxetable}{ccc}[ht]
\tablenum{1}
\tablecaption{Comparison between CLU and IPHAS
\label{tab:clu_iphas}}
\setcounter{table}{1}
\setlength{\tabcolsep}{0.8pt}
\tablehead{
\colhead{Property} &
\colhead{CLU} & 
\colhead{IPHAS}
}
\startdata
Location & San Diego, USA & La Palma, Spain \\
Observatory & Palomar & Roque de los Muchachos \\
Telescope & Samuel Oschin & INT \\
Aperture & 1.2\,m & 2.5\,m \\
Instrument & PTF Survey Camera & Wide Field Camera \\
Pixel Scale & 1.01$''\, \rm pixel^{-1}$ & 0.33$''\, \rm pixel^{-1}$ \\
Filters & H$\alpha$1, H$\alpha$2, H$\alpha$3, H$\alpha$4 & $r',\ i', \ \rm H\alpha$ \\
\multirow{4}{*}{H$\alpha \ \lambda$\,(\AA)} & 6563.7 & \multirow{4}{*}{6568} \\
 & 6642.7\\ 
 & 6717.6 \\ 
 & 6808.6 \\
\multirow{4}{*}{H$\alpha \; \Delta \lambda$\,(\AA)} & 84.6 & \multirow{4}{*}{95} \\
 & 85.5 \\ 
 & 95.1 \\ 
 & 98.7 \\
Exposure Time & 60\,s & 120\,s \\
Magnitudes & AB & Vega \\
Detection Limit & $19.1 \pm 0.5$ & $20.3 \pm 0.3$ \\
Longitude Range & & $29^\circ < l < 215^\circ$ \\
Latitude Range & $ \rm Decl. > -20^\circ$ & $-5^\circ < b < +5^\circ$ \\
Observing Period &  2009 -- 2017/03 & 2003/08 -- 2012/11
\enddata
\tablecomments{
Col. (1): Properties of interest. ``H$\alpha \; \lambda$'' denotes the central wavelength of the H$\alpha$ filters, ``H$\alpha \; \Delta \lambda$'' indicates the full width at half-maximum, and ``Detection Limit'' presents the mean and standard deviations of 5$\sigma$ detection limits in magnitudes.
Col. (2): Relevant values for the CLU survey retrieved from \citet{law2009,rau2009,cook2019} and Cook et al. (submitted). Data for the four H$\alpha$ filters are listed respectively from top to bottom in relevant sections for ``H$\alpha \lambda$'', ``H$\alpha \; \Delta \lambda$'', and ``Detection Limit''. Additionally, although observations using the H$\alpha$3 and H$\alpha$4 filters cover the sky above $-20^\circ$ declination, the Galactic plane ($|b| \lesssim 3^\circ$) is avoided \citep[see][Figures~2 and 3]{cook2019}.
Col. (3): Relevant values for IPHAS retrieved from \citet{drew2005,gonzalez-solares2008,barentsen2014}.}
\end{deluxetable}

\section{Recovering Planetary Nebulae} \label{sec:rediscov}
We proceed to validate the technical feasibility and quantify the effectiveness of systematic PNe detection within CLU-H$\alpha$ images via visual inspections of a catalog of PN candidates observed in the IPHAS survey. The three primary reasons for referring to this catalog are: (1) it stands as the largest H$\alpha$ survey in the northern Galactic plane; (2) it has undergone systematic study for PNe detection, contributing both methodologically and in terms of source inclusion to the HASH database; and, (3) it provides deeper data compared to the CLU survey, ensuring sufficiently reliable validations for our visual recovery. As detailed in the following sections, we scrutinize corresponding CLU-H$\alpha$ images to determine if the object has been satisfactorily detected by CLU, and then conduct a statistical analysis.

\subsection{The INT Photometric H$\alpha$ Survey}
The IPHAS is a comprehensive photometric survey of the northern Galactic plane covering Galactic latitudes of $|b|<5^\circ$ and longitudes of $29^\circ < l < 215^\circ$ \citep[for details, see][]{drew2005,barentsen2014}. Utilizing the 2.5\,m INT at La Palma in the Canary Islands of Spain with a Field of View (FoV) of $34' \times 34'$, the survey employs photometric filters including the Sloan $r$ (centered at 6240\,\AA, 30\,s exposure), the Sloan $i$ (7743\,\AA, 10\,s), and the H$\alpha$ (6568\,\AA, 120\,s) \citep{drew2005}. The coverage of H$\alpha$ point sources is down to a magnitude of 20.3 at the 5-$\sigma$ level \citep[Vega system; see][]{barentsen2014} against 19.1 with CLU-H$\alpha$1 stacked images (\citealp{cook2019}; 2024, submitted). The depth of IPHAS is further coupled with the sensitivity for extended emission down to an H$\alpha$ surface brightness limit of $2.5 \times 10^{-16}\, \rm erg\,cm^{-2}\,s^{-1}\,arcsec^{-2}$ (against $1.6 \times 10^{-14}\rm \, erg~s^{-1}~cm^{-2}$ with CLU). Such high sensitivity of IPHAS facilitates its detection of emitting nebulae and their morphological structures in the more obscured and crowded regions of the Galactic plane.

In Table~\ref{tab:clu_iphas}, a detailed comparison between IPHAS and CLU is presented. Evidently, the IPHAS is a deeper survey than CLU. However, the advantage of the latter is its much wider sky coverage, enabling exploration of the yet unexplored parts of the sky.

\begin{figure}[ht]
    \centering
    \includegraphics[width=\linewidth]{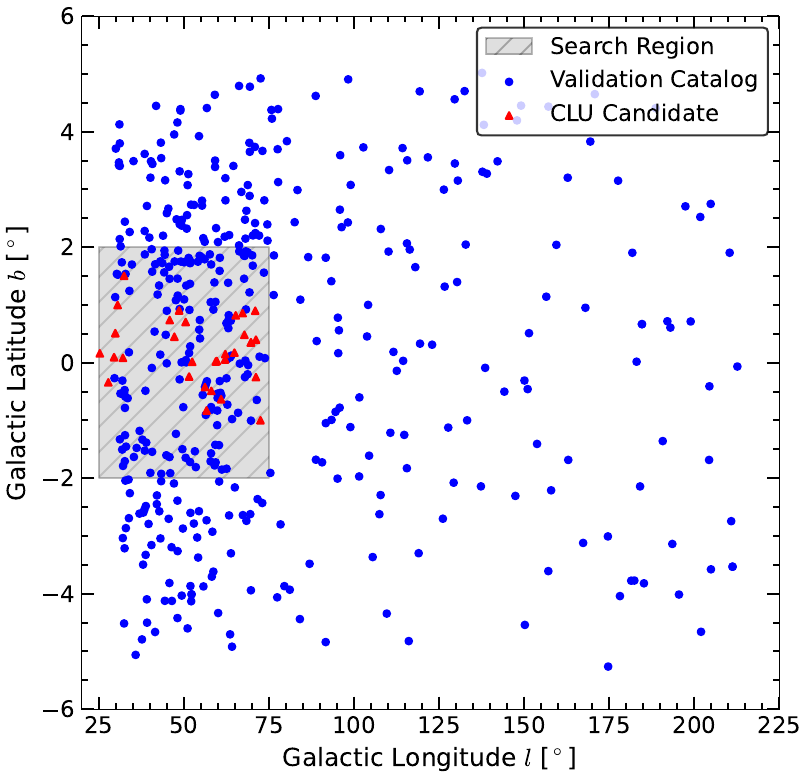}
    \caption{The sky locations in Galactic coordinates. The blue dots represent the planetary nebula candidates in the validation catalog. The red dots represent candidates from our preliminary search on CLU which is explained in Section~\ref{sec:preliminary}. The shaded region depicts the sky region of the preliminary search.}
    \label{fig:newpn_galactic}
\end{figure}
\subsection{The Validation Catalog} \label{sec:validation_catalog}
IPHAS provides imaging data for a plethora of H$\alpha$ emitting sources within which various studies have reported their discovering of PN candidates or even confirmed PNe with spectroscopic analysis \citep[e.g.,][]{corradi2005,viironen2009a,viironen2009b,sabin2010,sabin2014}. To construct the validation catalog, we retrieved a total of 441 labeled Galactic PNe detectable in IPHAS from the HASH database\footnote{\url{http://202.189.117.101:8999/gpne/}} (data retrieved in October 2023; \citealt{parker2016}).

Three searches contribute significantly to the catalog. The validation catalog encompasses 132 PNe originally observed by \citet{acker1992} but subsequently detected in IPHAS in succeeding works. As part of a systematic effort leveraging the IPHAS data, \citet{viironen2009a} and \citet{viironen2009b} discovered 70 sources using a semi-automated methodology that involved data cleaning primarily with cuts on IPHAS colors and also source selection based on both IPHAS and Two Micron All Sky Survey \citep[2MASS,][]{skrutskie2006} color indices. After color selection, all remaining objects were visually classified in their IPHAS images. A more recent work by \citet{sabin2014} contributed 105 additional extended sources through visual inspection of IPHAS H$\alpha-r$ difference map mosaics on two scales of binned data: initially by $5 \times 5$ pixels ($\sim 1.7''\,\mathrm{pixel}^{-1}$), and then by $15 \times 15$ pixels ($\sim 5''\,\mathrm{pixel}^{-1}$). This coarser binning aided in resolving low-surface-brightness objects effectively, reaching the detection limit of IPHAS as outlined by \citet{sabin2010}. For information on other references of the validation catalog, see Column~(13) of Table~\ref{tab:recover_iphas} in Appendix~\ref{app:iphas_result}.

In the HASH database, sources are documented with their name, coordinates in both Equatorial and Galactic systems, diameters of major ($D_\mathrm{maj}$) and minor ($D_\mathrm{min}$) axes, main and subsidiary morphology, position angle, multi-band fluxes, PN class, and reference catalog or article. The spatial distribution of all sources in the validation catalog is visualized in Galactic coordinates in Figure~\ref{fig:newpn_galactic}.

In HASH terminology, morphologies and classes adhere to a convention initiated by MASH \citep{parker2006_catalog}. Primary morphological classifications ``E; R; B; I; A; S'' denote Elliptical, Round, Bipolar, Irregular, Asymmetric, and quasi-Stellar (unresolved or barely resolved) PNe, respectively. Sub-morphology classifiers ``a; m; p; r; s'' are employed for ``a''symmetries, ``m''ultiple shells or external structures, ``p''oint symmetry, well-defined ``r''ing structures or annuli, and resolved, internal ``s''tructure, respectively. The PN class in terms of ``T/L/P'' denote the quality of sources during the identification process \citep[e.g.,][]{parker2005,parker2022}:
\begin{enumerate}[label={}]
    \item True ``T'': spectroscopically and morphologically well-defined PNe across multi-wavelengths, and sometimes with an obvious blue CSPN;
    \item Likely ``L'': not completely conclusive identification spectroscopically or morphologically;
    \item Possible ``P'': non-conclusive identification. The quality of spectroscopic or morphological data can resolve the source but cannot rule out contaminants like H\,{\small II} regions, etc.
\end{enumerate}
Out of the 441 sources in the validation catalog, 317 are classified as T, 49 as L, and 75 as P PNe, respectively.

\begin{figure*}[ht]
    \centering
    \includegraphics[width=0.78\linewidth]{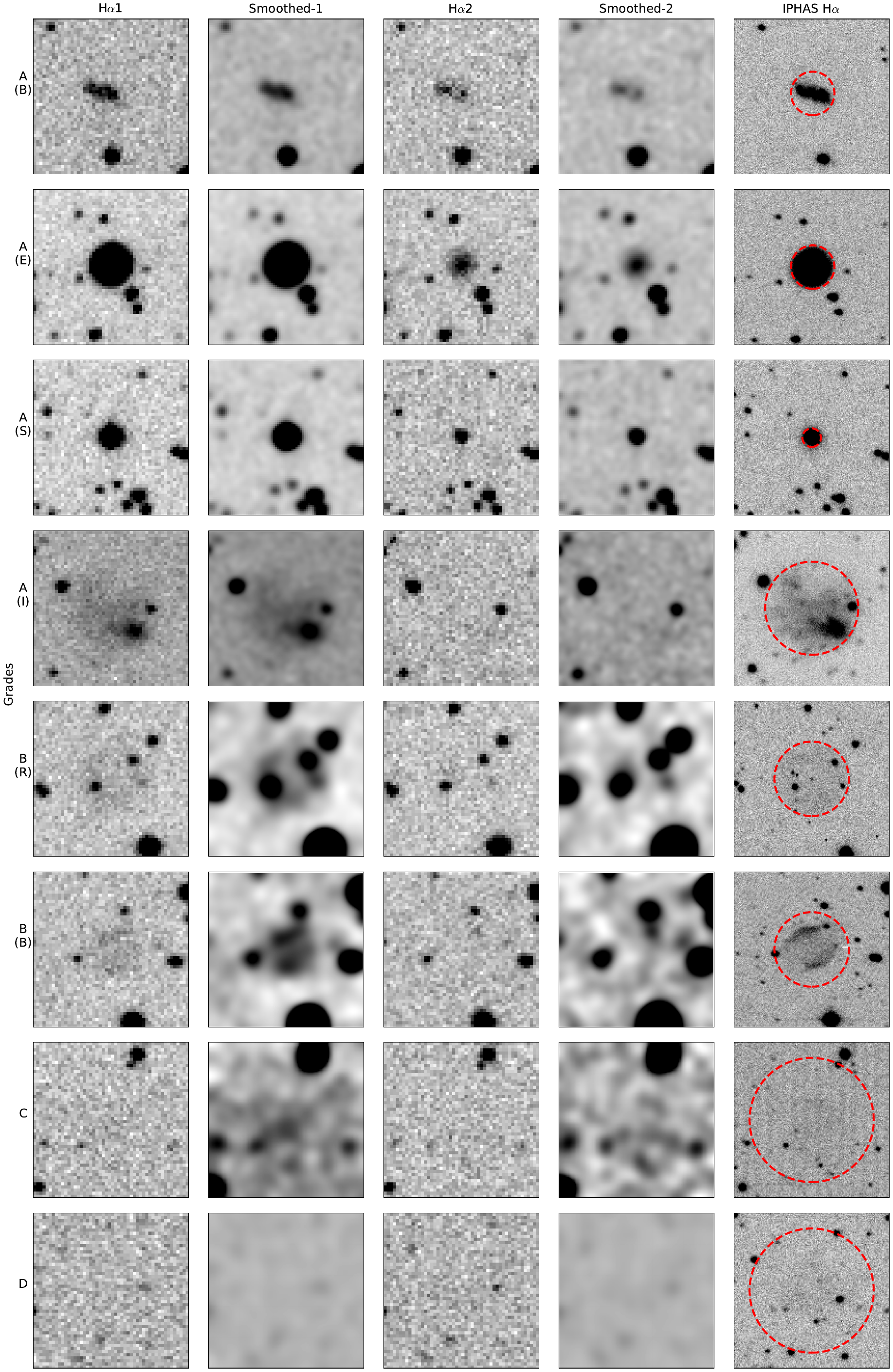}
    \caption{Example images of H$\alpha$ emission-line sources from the four PN classification grades from A to D (in the decreasing order of the confidence, c.f. Section~\ref{sec:vis_rediscov}). The five columns show images from the first H$\alpha$ band, smoothed H$\alpha$1, the second H$\alpha$ band, smoothed H$\alpha$2, and the respective IPHAS H$\alpha$ images. The red circles indicate the nebular regions. For visualization, the smoothing value is one pixel for grade~A sources, and two pixels for grades~B, C, and D sources. All images have a FoV of 50$''$. 
    \textbf{Row~1}: Images of a bipolar grade~A source (Te\,10, HASH ID 4372). 
    \textbf{Row~2}: An elliptical A source (K\,3-90, 653). 
    \textbf{Row~3}: A quasi-stellar A source (K\,3-69, 705). 
    \textbf{Row~4}:An irregular A source (IPHASX\,J193740.4$+$203547, 10061).
    \textbf{Row~5}: A round B source (IPHASX\,J013108.9$+$612258, 8450). 
    \textbf{Row~6}: A bipolar B source (IPHASX\,J194301.3$+$215424, 8566). 
    \textbf{Row~7}: A grade~C source (IPHASX\,J184030.1$-$003822, 8478) which is difficult to identify without referring to the IPHAS images showing the location of the emissions. \textbf{Row~8}: An undetected grade~D source (IPHASX\,J185225.0$+$080843, 15651).}
    \label{fig:grade_eg}
\end{figure*}

\subsection{Visual Inspection} \label{sec:vis_rediscov}
We conduct systematic visual inspections of images from CLU-H$\alpha$ centered on objects in the validation catalog. We note here that our goal is to determine the feasibility of finding PNe in the CLU survey, and that additional discovery methodologies for new PNe will be explored later in this paper (see Section~\ref{sec:preliminary}) and in future efforts. Since no spectroscopic follow-up is conducted, we refrain from adopting the ``T-L-P'' system used by HASH to register PNe status. Instead, we assign each source a classification grade based on both the quality of our visual detection of nebular emission and the confirmation of its morphology documented in previous catalogs:
\begin{enumerate}[label=(\Alph*)]
    \item Obvious nebular signal showing clear morphology, matching the known information.
    \item Clear detection of some part of the nebular emission with confident verification of the morphology. Additional visualizations (e.g., Gaussian smoothing, rebinning) and careful inspection are needed.
    \item Possible detection of only part of the nebular emission where the morphology cannot be robustly verified even when referring to the validation catalog.
    \item No detection of the nebula or any associated morphology.
\end{enumerate}
The grading system is based on the primary emission of PNe in H$\alpha$ (CLU-H$\alpha$1). For grade~A sources, the recovery is straightforward. For grade~B and C sources, we provide the reasons for such grading for each source. For grade~D sources, where the source is undetected regardless of fine-tuning the image visualization, we explain the reason of detection failures for each specific source. Sources with grade A or B are considered readily recovered. Grade~C sources are only marginally recovered.

Our procedure starts by examining CLU-Ha imaging at the coordinates of each validation object, where we adjust the FOV to approximately 1.5-2 times the major diameter of the nebula. If we can directly observe the source or, after contrast adjustments, discern a nebulosity and its documented morphology, the source is classified as grade~A. If confirmation is challenging, we apply Gaussian smoothing and designate the source as grade~B if the whole nebulosity or part of its morphology becomes visible. For sources that remain undetected, we refer to spectra, notes, and multi-band images in the HASH database, and adjust the image visualization parameters (image center, viewing contrast, $\sigma$ of Gaussian smoother, etc.) accordingly. Detected sources are in grade~C, while those undetected remain in grade~D. Examples are shown for the four respective grades in Figure~\ref{fig:grade_eg}.

\begin{figure}[t!]
    \centering
    \includegraphics[width=\linewidth]{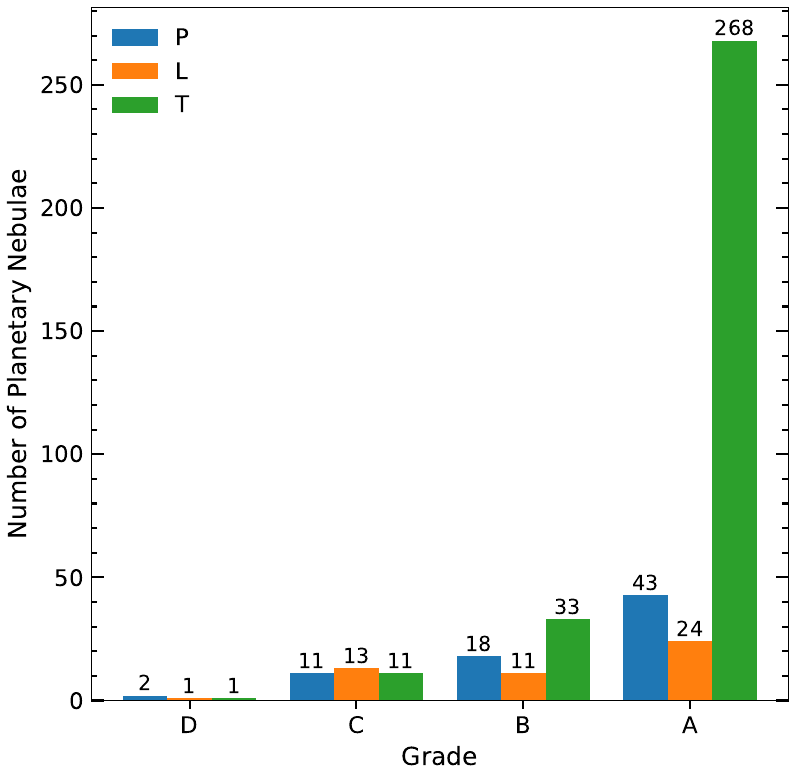}
    \caption{Histogram of planetary nebula candidates in the validation catalog with respect to different grades (A, B, C, and D). Within each grade, sources are divided according to their classes of True, Likely, and Possible.}
    \label{fig:grade_hist}
\end{figure}

Including five sources with unusable, masked CLU images that have flat photon distribution, we present the full result of the recovery in Table~\ref{tab:recover_iphas} (Appendix~\ref{app:iphas_result}). Overall, out of 441 validation sources, we have recovered 335~A (76\%), 62~B (14\%), and 35~C (8\%) sources, and labelled 4~D sources (0.9\%). Thus, the majority of sources (90\%) fall in grades~A and B as readily recovered cases. Including grade~C, we have in total 98\% validation sources successfully recovered, which demonstrates the ability of CLU-H$\alpha$ to reproduce previous discoveries, implying its viability for further research. A histogram of classification grades is shown in Figure~\ref{fig:grade_hist}. In PN classes, 95\% of True, 71\% of Likely, and 81\% of Possible sources are readily recovered, implying the consistency of high recovery rate across all classes. The fact that 85\% ``T'' sources are labelled as grades~A and B suggests that most of the bona fide PNe are easily recovered by CLU. For the four grade~D sources:
\begin{enumerate}[label={}]
    \item IPHASX J061602.5$+$175920 (class ``P''): It is a very large nebula ($D_\mathrm{maj} = 274''$) of very low surface brightness. CLU documented three single epoch images in the relevant sky region, from which a stacked image was created, but no nebular emission is detected. Two of the CLU exposures have masked regions or bad pixels in the nebular region, diminishing the quality of the ``stacked image'' to the same as the third single epoch unmasked image. Additionally, several bright point sources in the FoV of the unmasked epoch further restrain our visualization techniques to improve the contrast to reveal the emission region. Nevertheless, the source itself requires further observations for its nature to be determined.
    
    \item IPHASX J185225.0$+$080843 (class ``T'', reported by \citealt{sabin2014}): The source is very faint with low signal-to-noise ratio (S/N) in the IPHAS image. The relevant sky region was only observed once in the CLU-H$\alpha$ imaging data. This source is selected as an example to demonstrate our classification grades in Figure~\ref{fig:grade_eg}.

    \item Pa J1906.9$+$0413 (class ``L''): It was originally discovered by the Deep Sky Hunters community \citep{kronberger2006,jacoby2010}. Similar to the previous source, it has very weak IPHAS detection and only one CLU exposure. Nevertheless, it is well-resolved in the Wide-field Infrared Survey Explorer \citep[WISE;][]{wright2010} bands, implying possible obscuration along the line of sight.
    
    \item IPHASX J194745.5$+$270150 (class ``P''): It is a large ($D_\mathrm{maj} = 207''$), very faint, elliptical nebula observed four times by CLU. However, three of the four exposures are masked. The source also has an extremely flat spectral continuum around H$\alpha$, which make it hard to determine whether a real emitter with little H$\alpha$ excess or the background with no detection occupies the image.
\end{enumerate}

\begin{figure}[tb]
    \centering
    \includegraphics[width=\linewidth]{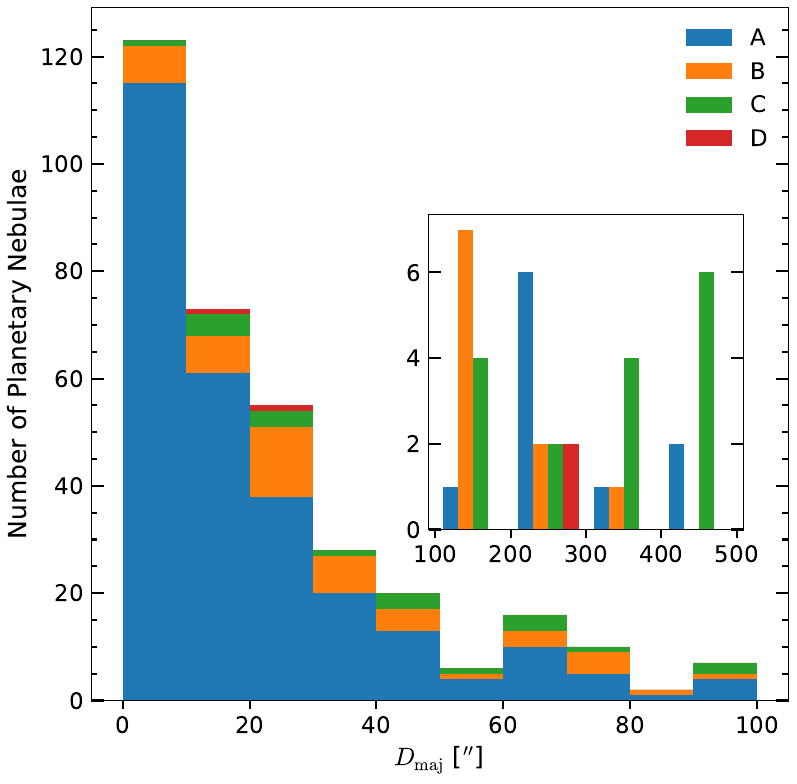}
    \caption{Histogram of planetary nebula candidates in the validation catalog with respect to different major diameters ($D_\mathrm{maj}$). Sources are labeled according to their grades of A, B, C, and D. In the figure, smaller nebulae are usually labeled with a higher grade ranking.}
    \label{fig:hist_d}
\end{figure}
\begin{figure}[ht]
    \centering
    \includegraphics[width=\linewidth]{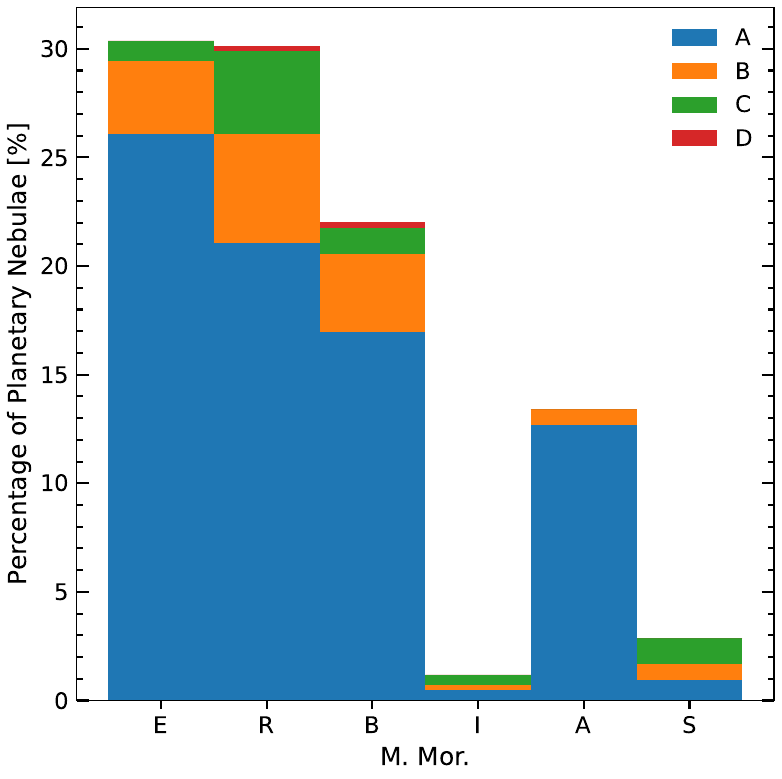}
    \caption{Histogram of planetary nebula candidates in the validation catalog with respect to different main morphology (M. Mor.), where E, R, B, I, A, and S stands for the M. Mor. of Elliptical, Round, Bipolar, Irregular, Asymmetric, and quasi-Stellar (unresolved or barely resolved), respectively. Sources are divided according to their grades of A, B, C, and D.}
    \label{fig:hist_mor}
\end{figure}
\subsection{Statistics on the Recovery}
We compute Spearman's rank correlation coefficient between the classification grades used in our study and the HASH PN classes, yielding $\rho = 0.3297$. A subsequent $T$-test returns a p-value of $p < 0.001$, which indicates a weak correlation between grades and classes. Better PN class means more prominent source and clearer morphology, which increases the chance of our visual detection, hence a higher grade.

Additionally, the apparent major diameter ($D_\mathrm{maj}$) of the nebula may influence detection. Excluding sources without reported diameters in the validation catalog, we present a histogram of $D_\mathrm{maj}$ in Figure~\ref{fig:hist_d}. With a Spearman's correlation coefficient $\rho = -0.4277$ between grades and $D_\mathrm{maj}$ and $p < 0.001$ in a subsequent $T$-test, we observe a weak-to-moderate negative correlation between grades and $D_\mathrm{maj}$ of the nebulae. A possible explanation would be that lots of larger nebulae may have lower surface brightness making them more difficult to detect.

The intricate and diverse morphology of PNe also serves as a vital classifier. Studies have suggested that the morphology of PNe can be indicative of the properties of their progenitor stars, mass ejection mechanisms, and multiple shaping processes involving external torques from close or merging binary companions, or the emergence of magnetic fields embedded in dense out-flowing stellar winds \citep[e.g.,][]{balick2002,parker2006,miszalski2009}. We present a histogram of main morphology categories (M. Mor.) in Figure~\ref{fig:hist_mor}, and notice the absolute dominance of those PNe with more symmetrical or barely resolvable quasi-stellar shapes. However, we are not able to make a conclusive claim, because a homogeneous sample of Galactic PNe would be needed. Given that IPHAS and SHS are deeper than CLU, our classification grades might be more affected by the resolvability of CLU images than the morphology of the sources.

\begin{figure*}[ht]
    \centering
    \includegraphics[width=\linewidth]{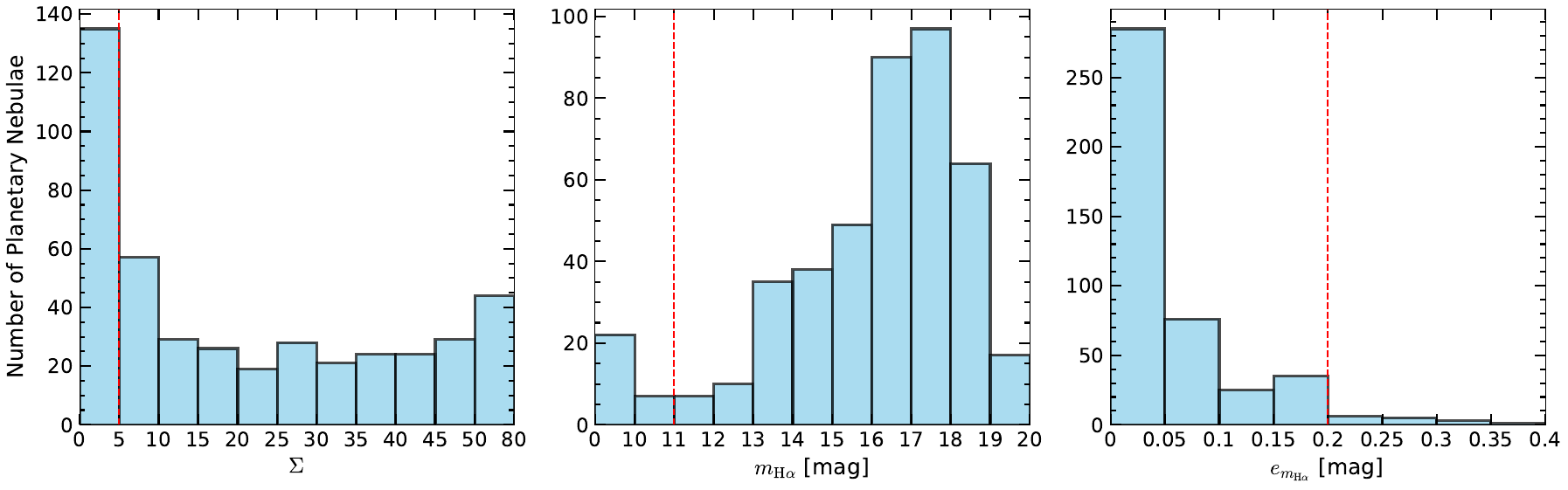}
    \caption{Histogram of planetary nebula candidates in the validation catalog with respect to different color sigma $\Sigma$, H$\alpha$ ``On'' band magnitude $m_\mathrm{H\alpha}$, and the error of the ``On'' magnitude ($e_{m_\mathrm{H\alpha}}$). The red dashed line depicts the filtering criteria in the catalog search.}
    \label{fig:hist_filterpara}
\end{figure*}
\begin{figure}[ht]
    \centering
    \includegraphics[width=\linewidth]{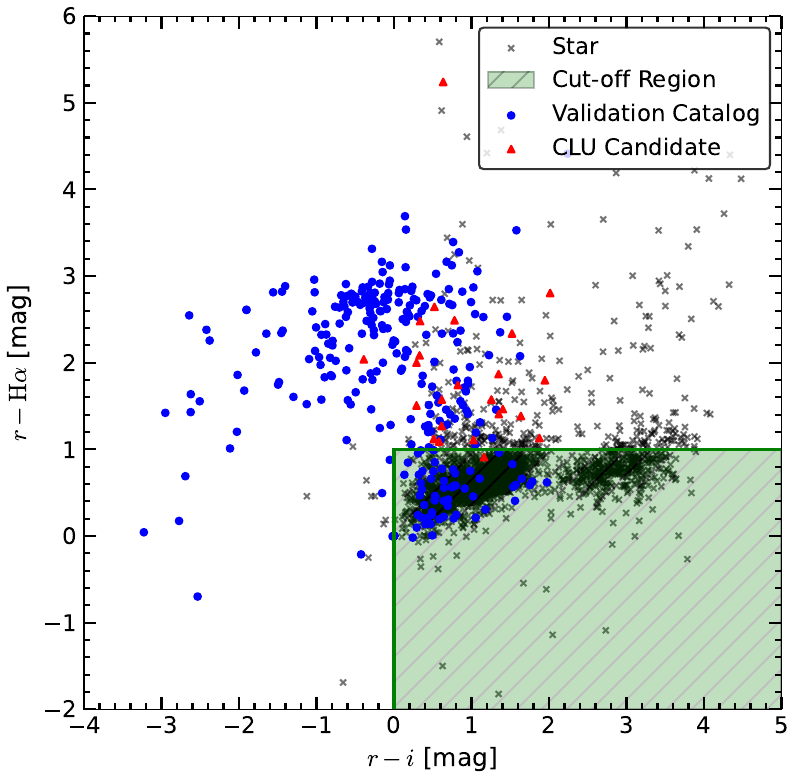}
    \caption{Color-color diagram depicting the ($r - \mathrm{H}\alpha$) versus ($r - i$) scatter plot of labelled stars from contaminant removal, planetary nebula candidates in the validation catalog, and candidates from our preliminary search on CLU. The green shaded region represents the region to be cut off through a rough preliminary color cut. All colors in the figure are \citet{kron1980} magnitudes.}
    \label{fig:color_cut}
\end{figure}
\begin{figure*}[ht]
    \centering
    \includegraphics[width=0.7\linewidth]{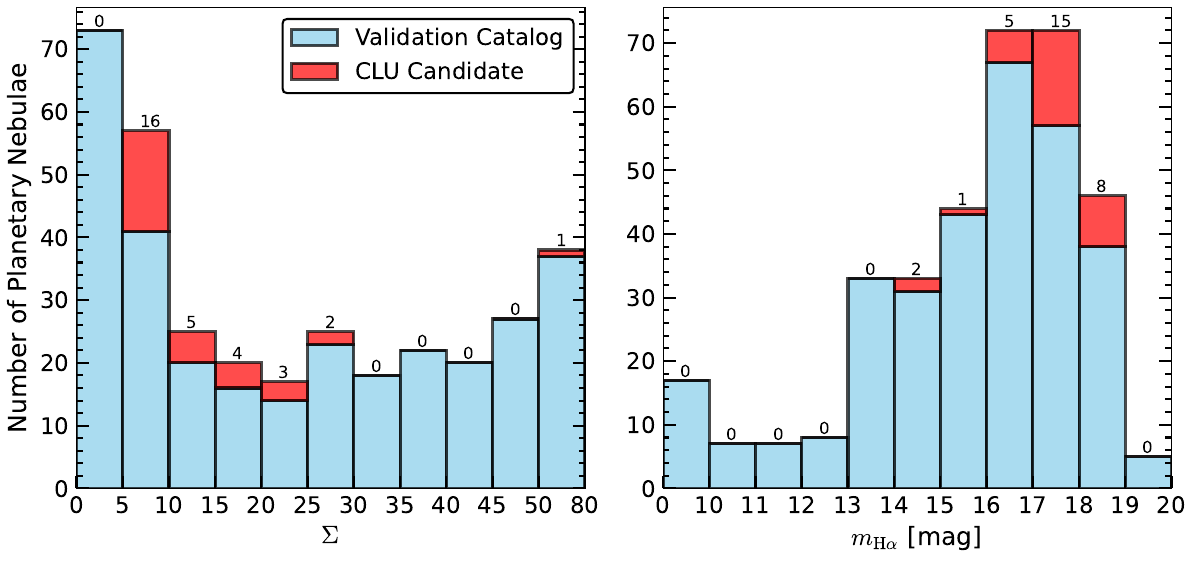}
    \caption{Histogram of True planetary nebula candidates in the validation catalog and PN candidates from our preliminary search on CLU, with respect to different color sigma ($\Sigma$; left panel) and H$\alpha$1 magnitude ($m_\mathrm{H\alpha}$; right panel). The blue bars represent sources in the validation catalog, while the red bars represent our candidates. The numbers atop of each bar represent the counts for CLU PN candidates.}
    \label{fig:newpn_properties}
\end{figure*}

\section{Preliminary Search on the CLU Survey} \label{sec:preliminary}
Previous analyses have shown that known PNe can be detected from CLU images effectively, whereas in this section, we briefly explore the potential of discovering new PNe in the CLU database. We start with a catalog search for strong emission-line sources in the CLU dataset with $|b| < \pm 2^\circ$ and $l = 25^\circ$ to $75^\circ$ to identify PN candidates. Subsequent spectroscopic follow-up for six candidates are presented. We recognize that our preliminary search does not aim to offer a systematic and complete CLU PNe list for the selected region, because such discoveries of new PNe based on their nebular emission usually necessitates the combination of several different techniques due to the complex characteristics of PNe including apparent size, morphology, integrated flux, excitation state, CSPN magnitude, and evolutionary state, unless the search focuses on identifying clear nebular regions around known CSPNe \citep[e.g., see the HASH approach in][Section~12.1]{parker2022}. However, the exercise we present in this section is sufficient to demonstrate the significant promise of discovering new PNe in CLU.

\subsection{The Catalog Search} \label{sec:cat_search}
We utilize the CLU forced photometry source catalog called CLU-FPstack described in Zhang et al. (2024; submitted), where H$\alpha$ fluxes are extracted from the coordinates of Pan-STARRS Data Release 1\citep[PS1,][]{chambers2016} objects regardless of the detection significance in the CLU images. We search CLU-FPstack for PNe, where parameter cut-offs are applied to produce robust candidates. The source catalog search selects candidates primarily based on the ``color sigma'' \citep[$\Sigma$;][]{cook2019}, which quantifies the significance that the narrow-band color excess is in one filter above the combined noise of this filter and its adjacent filter. We also ensure the candidates have robust photometry and subsequently robust $\Sigma$ measurements by applying more cuts on parameters as outlined in the following section. After the selection, we remove contaminants based on their color features and morphology to finally present a list of 31 CLU candidates from the preliminary search.

\subsubsection{Source Selection}
Widely used for identifying H$\alpha$ emission-line sources \citep[e.g.,][]{bunker1995,pascual2001,fujita2003,sobral2009,sobral2012,ly2011,lee2012,stroe2015,cook2019}, $\Sigma$ is based on the ``On-Off'' color excess. Our search for PN candidates focuses on the excess of H$\alpha$ emission to the continuum, with H$\alpha$1 serving as the ``On'' band and H$\alpha$2 as the ``Off'' band, which gives:
\begin{equation}
    \Sigma = \frac{c_\mathrm{H\alpha1} - c_\mathrm{H\alpha2}}{\delta},
\end{equation}
where $c_{H\alpha1, 2}$ are the counts for H$\alpha$1 and H$\alpha$2 bands respectively, and $\delta$ is the combined sky fluctuations of both filters, of which detailed explanations are provided in Section~3.2 of \citet{cook2019}. Using a fixed photometric aperture with 5$''$ of diameter, we require $\Sigma > 5.0$ to indicate a positive detection (\citealt{cook2019}, 2024; submitted).

Since the source catalog fluxes from forced photometry are based on the locations of the PS1 objects, the offset distance from the PS1 source to the actual CLU source adversely affects the robustness of photometric property calculations. By imposing an upper limit on the distance to the source detected in the stacked image ($dist2stack < 1.5''$), we filter sources whose optical emission centers viewed in PS1 are farther from their positions of H$\alpha$ emission in CLU stacked images. Additionally, we set a minimum H$\alpha$ ``On'' band magnitude of 11 ($m_\mathrm{H\alpha} > 11$) to remove saturated sources with poor flux measurements. We further adopt an upper limit for the error of the ``On'' magnitude to 0.2 ($e_{m_\mathrm{H\alpha}} < 0.2$) for $5\sigma$ detections.

Prior to applying the cuts to real data, it is useful to estimate its success rate, for which we refer to the previously defined PNe validation catalog (c.f. Section~\ref{sec:validation_catalog}). Since the region for our preliminary search is within the coverage of the validation catalog as depicted in Figure~\ref{fig:newpn_galactic}, we can test the cuts on the validation catalog and safely use the results for inferences in our preliminary search region. We apply the cuts to the validation catalog, and notice that 67.9\% of sources satisfy $\Sigma > 5.0$, 93.4\% have $m_\mathrm{H\alpha} > 11$, and 96.6\% get $e_{m_\mathrm{H\alpha}} < 0.2$. We present a histogram in Figure~\ref{fig:hist_filterpara} to illustrate the cut on the distribution of parameters. We notice that $\sim 135$ sources with $\Sigma<5$ are cut off in the left panel, which take up to $\sim 30\%$ of the validation catalog. For cuts on $m_\mathrm{H\alpha}$ and $e_{m_\mathrm{H\alpha}}$, a small portion is also discarded. In total, 66.1\% sources satisfy all the criteria. Therefore, by applying our cuts, we could potentially cut off $\sim 1/3$ actual PNe through the source catalog search. 

In general, we rule out sources with bad photometry or marginal significance by the cuts, but also suffer from missing PNe with less excess in H$\alpha$ or with subtle signals. Implementing the search criteria to CLU-FPstack within $|b| < 2^\circ$ and $25^\circ < l < 75^\circ$, we are able to select $\sim 10,000$ emission-line candidates.

\subsubsection{Contaminant Removal} \label{sec:color_cuts}
Our straightforward selection criteria add all the H$\alpha$ emission-line sources into the candidate list which may suffer from contamination by various interlopers, such as stars with strong emission lines, H\,{\small II} regions, Wolf-Rayet shells, supernova remnants (SNRs), nearby emission-line galaxies, and artifacts in the CLU images including cosmic rays, hot pixels, satellite streaks, and image ghosts that are not removed during the stacking process. General removal methodology is complicated, involving information from spectroscopy (c.f. Section~\ref{sec:caveats}). For the preliminary search, we rely on simple color cuts and a few visual inspections to conduct the removal.

Star removal always remains challenging due to its resemblance to compact PNe in their point-like appearances and photometric properties investigated during our detection. Here, we refer to a traditional method of color cuts on two indices \citep[e.g.,][]{viironen2009a,viironen2009b}. We conduct the H$\alpha$-PS1 cut-off criterion by removing sources with both $r - i > 0$ and $r - \mathrm{H}\alpha < 1$ as demonstrated in the color-color diagram of ($r - \mathrm{H}\alpha$) versus ($r - i$) in Figure~\ref{fig:color_cut}. The 3400 example stars in the plot are manually classified by us with visual scrutiny on the preliminary search region for testing purposes. Stars in the plot exhibit two distinct clustering trends, indicating the presence of two stellar populations in the CLU survey. For the effectiveness of the cuts, we use the validation catalog constructed in the recovery (Section~\ref{sec:validation_catalog}) again to calculate that this criterion retains 96\% of their documented PN while eliminating 88\% of example stars. Applying the color-color cut-off to all $10^4$ emission-line candidates, we are able to remove $48\%$ of them as stars.

For the rest of the candidates, we visually label them purely based on the clearness of their nebular morphology and remove the other contaminants mainly consist of artifacts that have singular morphology only existing in one band (48\%). During the visual scrutiny, we also refer to images and previous classifications, if any, documented by PS1, Sloan Digital Sky Survey III \citep[SDSS-III;][]{eisenstein2011}, Set of Identifications, Measurements and Bibliography for Astronomical Data \citep[SIMBAD;][]{wenger2000}, and HASH, when necessary. With these additional information, if certain sources are firmly considered as a type of contaminant, they are removed by us as well. In the end, $\sim 400$ sources are identified as other astrophysical objects (e.g., Milky Way H\,{\small II} regions, galaxies, red stars, etc.).

\subsection{Planetary Nebula Candidates} \label{sec:new_pne}
After candidate selection and contaminant removal, we finally uncover 31 PN candidates in $|b| < 2^\circ$ and $25^\circ < l < 75^\circ$ of the CLU-FPstack catalog. We illustrate the spatial distribution of the sources in Galactic coordinates in Figure~\ref{fig:newpn_galactic}, and summarize the notes and respective coordinates of those PN candidates in Table~\ref{tab:clupne}.

During the visual inspection, we distinguish the PN candidates and label them with the classification grades (c.f. Section~\ref{sec:vis_rediscov}) based on the visibility of its morphology, as presented in Column~(6) of the table. In brief, grade~A sources exhibit very clear and distinct morphological features by eye, whereas grade~B represents more ambiguous sources that need adjusting the visualization parameters of CLU images to make it visible. Since we are searching for PNe rather than recovering previous work, we are not able to discern the unclear C, needless to say D, sources without references. Among the candidates, 19 (61\%) are grade~A with clear morphology, and 12 are grade~B with only visible morphology after tuning the CLU image. Apart from the clearness of morphology, we also record the references and additional information on its goodness of morphology. We could infer from notes in Column~(7) that 12 (39\%) candidates are compact and also likely to be stars, and that 4 (13\%) are very diffuse and likely to be Galactic H\,{\small II} regions, leaving the rest 15 candidates that do not share similarities with possible contaminants. However, the 31 candidates have all passed our contaminant removal procedure, which means that all sources may be considered as PN candidates.

For references from previous work, 19 of our 31 candidates are reported in other studies initially as PN candidates as well \citep{parker2006,sanchezcontreras2008,viironen2009a,viironen2009b,sabin2014,froebrich2015,irabor2018,Kuhn21}, whereas the remaining 12 sources are newly identified. Within the 19 known sources, 10 of them are suspect PNe, 2 are actually the bright parts of discovered True PN documented in HASH, 3 are as likely to be PNe as alternative contaminants, 3 are not suggested to be PN by HASH, and 1 appears as a young stellar object (YSO) candidate on SIMBAD. Merely limiting to the new candidates, 5 of them have rather clear morphology (i.e., grade~A source), which have our highest confidence to be real PNe. Nevertheless, spectroscopic follow-up on our candidates will serve as the necessary final step for the confirmation of their identity, which we generally leave for future work.

In Figure~\ref{fig:newpn_properties}, we plot the distribution of $\Sigma$ and H$\alpha$1 magnitude for both True PNe in the validation catalog and candidates from our preliminary search. From the plot, we see effects of catalog search cuts to the property distribution of CLU candidate. In the left panel, we plot the distribution in $\Sigma$ space, where no CLU sources are detected with $\Sigma < 5$ as expected. Even though 73 of 317 True cataloged PNe have $\Sigma < 5$, the rest 77\% in the catalog still have a larger color excess, indicating that using the $\Sigma$-based catalog search can indeed find new True PNe. Such sources will exist in our preliminary search candidates and the future CLU PNe catalog. In the right panel, the distribution is demonstrated for the H$\alpha$ magnitude space, where no CLU candidates locate in $m_\mathrm{H\alpha} < 11$ or beyond the 19.1 magnitude limit of CLU. The peak of the CLU candidates from our preliminary catalog search are on the fainter end of the distribution, i.e., 1 magnitude dimmer than the validation sources, which indicates that our catalog search will likely find fainter candidates that are more likely to be omitted by visual inspections in previous work.


\subsection{Follow-up Spectroscopic Observations}
Detection in H$\alpha$ is usually not enough to classify an object as a PN and requires spectroscopic follow-up. We are in the process of obtaining spectroscopic data for several of the candidates for appropriate classification. Here, we present some initial results: low-resolution optical spectroscopic observations of six candidates (from Table \ref{tab:clupne}). The sample selection is performed based on the H$\alpha$ morphology of the object, its position in diagnostic diagrams (like Figure \ref{fig:color_cut}), and/or detection in other wavelengths, for example infrared (IR) detection in WISE. A more complete study will be presented in a future work.

The observations were performed using the DouBle SPectrograph (DBSP; \citealt{Oke82}) attached to the Cassegrain focus of the Palomar 200-inch Hale Telescope. We use the D55 dichroic, the 600 line~mm$^{-1}$ grating for the blue arm blazed at 3780~\AA, and the 316 line~mm$^{-1}$ grating for the red arm blazed at 7150~\AA. We use grating angles of 27$^{\circ}$17$'$ and 24$^{\circ}$38$'$ for the blue and red sides, respectively. With these setups and a slit width of 1.5$''$ long-slit, we achieve resolving powers of $R=1600$ in the blue arm and $R=1400$ in the red arm. This setup provide continuous spectral coverage across 2900--10800~\AA, with the division between blue and red arms typically occurring around 5650~\AA. The standard star Feige\,110 is used for flux calibration.

The standard PypeIt \citep{PypeIt_1, PypeIt_2} based optimal-reduction pipeline for DBSP spectra, DBSP\_DRP \citep{Roberson21}, was often seen to be masking bright and extended emission lines, rendering it unsuitable for PNe. Thus, we manually run PypeIt to perform the reduction. We follow the PypeIt recommendations for bright emission lines, and set the parameters \texttt{use\_2dmodel\_mask = False} and \texttt{no\_local\_sky = True}, to avoid masking and local subtraction of the lines. We perform a careful cosmic ray rejection on the two-dimensional spectra using \texttt{lacosmic} \citep{van-dokkum2001}, ensuring that none of the emission lines are rejected. We then perform manual trace identification on the two-dimensional spectrum. We use a trace width of six pixels in the spatial direction to obtain the one-dimensional spectrum. We make sure that the extended emission lines are not affected during background subtraction. Owing to the contamination of the one-dimensional spectra from residual cosmic rays, we closely refer to the two-dimensional spectra while identifying emission lines. For the purpose of this work, only the most prominent emission lines are identified. We discuss the spectral properties of the objects briefly in the following paragraphs. The spectra, along with the CLU images are provided in Appendix \ref{app:dbsp_spectra} (Figure \ref{fig:combined1}). 

\emph{CLU\,J183648.44$-$064442.8}: This object does not have any prior identification/reference. It is an extended source, with a major diameter ($D_\mathrm{maj}$) of $18''$. Interestingly, the apparent H$\alpha$ morphology resembles that of a bow shock, though sufficient data is not available in favor of or against this scenario. The spectrum is dominated by nebular emission lines; no stellar continuum is detected. The emission lines identified in the spectrum are the first three lines in the Balmer series (H$\alpha,~\beta,~\gamma$, with the last one being very weak) and the forbidden emission lines of [\ion{O}{2}]~3727,~3729~\AA, [\ion{N}{2}]~6548,~6583~\AA, [\ion{S}{2}]~6716,~6737~\AA, and [\ion{S}{3}]~9062,~9532~\AA. Interestingly, the spectrum lacks [\ion{O}{3}] lines. Such a feature is often attributed to young and very low excitation (VLE) PN. However, the simultaneous presence of the high excitation [\ion{S}{3}] lines is puzzling.  Strong [\ion{S}{3}] with only faint emission features in the blue are often seen in highly reddened PNe (see for example \citealt{Jacoby04,fragkou2018}). This may be a low excitation PN where [\ion{O}{3}] lines are intrinsically weak, which got further weakened (and, thus, undetected) with reddening. Using the line ratio of H$\alpha$ to H$\beta$ (Balmer Decrement, BD$_{\alpha\beta}$), and using ${E_{B-V}} \approx {\rm 1.9\log\left({\rm BD_{\alpha\beta}}/2.8\right)}$, we get ${E_{B-V}}$$\simeq$$0.94$, which is indeed quite high. Very low excitation (VLE) PNe usually are young and have high densities. For this object, however, we obtain a [\ion{S}{2}]~6716~\AA/6737~\AA\ line intensity ratio of $\sim$1.15, which, for reasonable temperatures, correspond to low electron densities $n_e$$\lesssim$$500~{\rm cm^{-3}}$ \citep{Osterbrock04}, contrary to our expectation. We further investigate the position of this object in the $\log({\rm H\alpha}/\text{[\ion{N}{2}]})$ -- $\log({\rm H\alpha}/\text{[\ion{S}{2}]})$ space \citep{frew2010}. We obtain $\log({\rm H\alpha}/\text{[\ion{N}{2}]})$$=$$0.046$ and $\log({\rm H\alpha}/\text{[\ion{S}{2}]})$$=$$0.175$. This places the object on the margin between SNRs and PNe. We note here that the neighborhood of this object has significant nebulosity. Several other localized emissions (of varying morphology) are also visible in the Pan-Starrs image. This object may not be PN but simply a denser part of a much larger ionized ISM/SNR. Further observations are needed to understand this candidate better.

\emph{CLU\,J184324.79$-$014319.2}: This is an unresolved point source in the CLU H$\alpha$ image. The spectrum shows a clear signature of a late-type star in the red, with H$\alpha$. Overall, the spectrum closely resembles that of a symbiotic system (SySt) or a cataclysmic variable (CV). The accreting white dwarf is possibly too faint to be detected in the blue spectrum. This object, thus, can be considered a false positive.

\emph{CLU\,J192710.49$+$162128.6}: This is also an unresolved point source. It has previously appeared in the list of IPHAS candidate PN in \citep{viironen2009b} and remains listed among ``New Candidates'' in the HASH catalog. Prior to this work, no spectroscopic observation of this object was available. Our spectrum lacks a lot of features except the Balmer series. An emission signature at the [\ion{O}{1}]~7774~\AA\ triplet is evident. However, it is unclear whether this is a true feature or a cosmic-ray artifact. The spectrum, overall, is not conclusive. Without the presence of high excitation lines like [\ion{O}{3}], the object remains a candidate for compact VLE PN. However, it can also be a false-positive, like an emission-line star.

\emph{CLU\,J194817.60$+$242116.7}: This is again an unresolved point source. It appears as a YSO candidate on SIMBAD from \citet{Kuhn21}. The spectrum, however, shows the signature of a late-type star in the red. Furthermore, it shows the Balmer and \ion{Ca}{2}~H,~K,~8498,~8542,~8662~\AA\ emission lines. These are typical features of a SySt or CV, making this object another false positive. 

\emph{CLU\,J194804.17$+$254848.6}: This is a compact but extended source in H$\alpha$, with a $D_\mathrm{maj}$ of $9''$ in the CLU image. The object appears among extended H$_2$ sources in \citet{froebrich2015} and as a PN candidate in \citet{jones2018} through near-IR K-band spectroscopy. The object also appears among ``New Candidates'' in HASH. The spectrum shows several emission lines, the prominent ones being: H$\alpha,~\beta$;~[\ion{O}{3}]~5007,~4959~\AA; [\ion{S}{2}]~6716,~6737~\AA; [\ion{N}{2}]~6548,~6583~\AA; [\ion{S}{3}]~9069,~9532~\AA; and [\ion{O}{1}]~3727,~3729~\AA. All of these lines are typical of PNe. The spectrum also appears to show a stellar continuum with absorption features. However, we suspect that this is from an unrelated field star. We obtain $\log({\rm H\alpha}/\text{[\ion{N}{2}]})$$=$$-0.064$ and $\log({\rm H\alpha}/\text{[\ion{S}{2}]})$$=$$0.23$. Similar to CLU\,J183648.44$-$064442.8, this places the object again very close to the margin between PNe and SNRs/H\,{\small II} regions in the diagnostic diagram. With [\ion{O}{3}] lines present, we perform a second test with the diagnostic space of $\log(\text{[\ion{O}{3}]~5007~\AA}/H_{\beta})$ -- $\log(\text{[\ion{N}{2}]~6584~\AA}/H_{\alpha})$ (Figure 5 in \citealt{frew2010}). We obtain $\log(\text{[\ion{O}{3}]~5007~\AA}/H_{\beta})$$=$$0.59$ and $\log(\text{[\ion{N}{2}]~6584~\AA}/H_{\alpha})$$=$$-0.05$. This is consistent with the object being PN, but the possibility of it being an SNR cannot be ruled out. We also obtain a [\ion{S}{2}]~6716~\AA/6737~\AA\ line intensity ratio of $\sim$1.2, which corresponds to $n_e$$\lesssim$$500~{\rm cm^{-3}}$. We obtain BD$_{\alpha\beta}$$=$10.03, which leads to a high reddening of $E_{B-V}$$=$$1.1$. Though further observations and analyses are needed to confirm the nature of this object, it stands as a good candidate for a True PN.

\emph{CLU\,J201756.61$+$335718.2}: This is another compact but extended H$\alpha$ source, with $D_\mathrm{maj} = 12''$ in the CLU image. A star at the location of the nebula is detected in PS1, which can be the central star. It is also a \emph{Gaia} source with $G$$=$$20.7$ and $G_{\rm BP}-G_{\rm RP}$$=$$0.74$, but without any reliable parallax information. The nebula is reported by the amateur community. It is currently listed as a Possible PN in HASH with a very low S/N spectrum, with a need for a higher-quality spectrum mentioned. The spectrum shows emission lines typical of PNe: H$\alpha,~\beta$;~[\ion{O}{3}]~5007,~4959~\AA, [\ion{S}{2}]~6716,~6737~\AA, [\ion{N}{2}]~6548,~6583~\AA, and [\ion{O}{1}]~3727,~3729~\AA. We perform the same diagnostics as CLU\,J194804.17$+$254848.6. We obtain $\log({\rm H\alpha}/\text{[\ion{N}{2}]})$$=$$0.033$ and $\log({\rm H\alpha}/\text{[\ion{S}{2}]})$$=$$0.594$. This is consistent with the object being a PN, but it is also close to the spaces for SNR\slash \ion{H}{2}. Additionally, we get $\log(\text{[\ion{O}{3}~5007~\AA]}/H_{\beta})$$=$$0.579$ and $\log(\text{[\ion{N}{2}]~6584~\AA}/H_{\alpha})$$=$$-0.149$. In this space, the object lies away from the crowd of either SNR or \ion{H}{2} regions, and appears more consistent with being a PN. We obtain a [\ion{S}{2}]~6716~\AA/6737~\AA\ line intensity ratio of $\sim$1.18, which again corresponds to $n_e$$\lesssim$$500~{\rm cm^{-3}}$. The object, however, does not appear to be as severely reddened. We obtain BD$_{\alpha\beta}$$=$4.28, which leads to a moderate reddening of $E_{B-V}$$=$$0.34$. Overall, the object stands as a strong candidate for a True PN. 

\movetabledown=60mm
\begin{rotatetable*}
\begin{deluxetable*}{cCCCCcl}
\tablenum{2}
\tablecaption{Thirty-One Planetary Nebula Candidates from the Preliminary Search
\label{tab:clupne}}
\setcounter{table}{2}
\setlength{\tabcolsep}{1pt}
\tablehead{
\colhead{Name} &
\colhead{R.A. (J2000)} & 
\colhead{Decl. (J2000)} & 
\colhead{$l$} & 
\colhead{$b$} & 
\colhead{Grade} &
\colhead{Notes}
\\
& 
\colhead{($^\circ$)} & 
\colhead{($^\circ$)} &
\colhead{($^\circ$)} & 
\colhead{($^\circ$)} &
}
\decimalcolnumbers
\startdata
CLU J183648.44$-$064442.8 & 279.20195 & -6.74537 & 25.26449 & 0.16175 & A & Diffuse, also possibly an H\,{\small II} region; clear morphology \\
CLU J184312.04$-$044522.6 & 280.80017 & -4.75628 & 27.76107 & -0.34024 & B & Red star or PN; visible morphology \\
CLU J184324.79$-$014319.2 & 280.85328 & -1.72201 & 30.48428 & 0.99928 & A & Faint \& compact; clear morphology  \\
CLU J184443.03$-$030355.8 & 281.17928 & -3.06551 & 29.43786 & 0.09585 & B & Masked in an epoch; inconclusive with a bright star in the FoV  \\
CLU J184932.49$-$004429.9 & 282.38538 & -0.74163 & 32.0557 & 0.08374 & A & Very diffuse, also possibly an H\,{\small II} region; clear morphology in FoV of 150$''$ \\
CLU J193948.71$+$203344.0 & 294.95297 & 20.56221 & 56.63976 & -0.82667 & B & Milky Way H\,{\small II} region or PN; visible morphology \\
CLU J194130.75$+$215650.2 & 295.37815 & 21.94729 & 58.03997 & -0.48734 & B & Compact, also possibly a star; visible morphology \\
CLU J194224.54$+$232009.7 & 295.60226 & 23.33603 & 59.34859 & 0.02151 & B & Compact, also possibly a star; visible morphology  \\
CLU J194302.30$+$233615.4 & 295.75959 & 23.60428 & 59.65308 & 0.02958 & B & Faint \& compact, also possibly a star; visible morphology \\
CLU J194816.04$+$254227.7 & 297.06684 & 25.70769 & 62.06862 & 0.05648 & A & Compact, also possibly a star; clear morphology \\
CLU J200514.13$+$322139.7 & 301.30889 & 32.36103 & 69.67817 & 0.35278 & B & PN or symbiotic star, also likely a jet component of IRAS\,20032+3212 \\
CLU J200617.44$+$334335.9 & 301.57267 & 33.72664 & 70.94905 & 0.89943 & A & Compact, also possibly a star; clear morphology \\ \hline
CLU J184355.71$-$023220.2 & 280.98211 & -2.53894 & 29.81633 & 0.51156 & A & Part of a True PN (R.A.$=280.9871$, Decl.$=-2.5356$, \citealt{parker2006}) \\
CLU J184505.23$+$001158.6 & 281.27181 & 0.19961 & 32.38552 & 1.50401 & A & None PN suspected in HASH \citep{viironen2009b} with low S/N spectrum \\
CLU J191233.22$+$114631.3 & 288.13844 & 11.77535 & 45.80034 & 0.73884 & A & Asymptotic giant branch star suspected in HASH \citep{viironen2009a} \\
CLU J191614.08$+$125202.2 & 289.05872 & 12.86728 & 47.18501 & 0.44987 & A & Studied by \citet{froebrich2015} and \citet{jones2018}; clear morphology \\
CLU J191713.48$+$141747.2 & 289.30615 & 14.29645 & 48.56155 & 0.9035 & B & Suspected PN in SIMBAD reported by \citet{irabor2018};  visible morphology \\
CLU J192140.59$+$155353.8 & 290.41919 & 15.89823 & 50.48032 & 0.70474 & A & Symbiotic Star or young PN in HASH \citep{sabin2014}; clear morphology \\
CLU J192710.49$+$162128.6 & 291.79372 & 16.35794 & 51.51049 & -0.24058 & A & Compact, reported by \citet{viironen2009b}; clear morphology \\
CLU J192802.95$+$171643.3 & 292.01229 & 17.27869 & 52.41955 & 0.01468 & B & None PN suspected in HASH \citep{viironen2009a} without spectrum \\
CLU J193719.79$+$202112.4 & 294.3325 & 20.35345 & 56.17254 & -0.42261 & A & Part of a True PN (R.A.$=294.3278$, Decl.$=20.3508$, \citealt{sabin2014}) \\
CLU J194804.17$+$254848.6 & 297.01736 & 25.81351 & 62.13741 & 0.14842 & A & Diffuse, reported by \citet{froebrich2015}; clear morphology \\
CLU J194817.60$+$242116.7 & 297.07333 & 24.35465 & 60.90422 & -0.63268 & A & Compact, young stellar object candidate in SIMBAD \citep{Kuhn21} \\
CLU J195229.51$+$284735.8 & 298.12297 & 28.79328 & 65.2025 & 0.81912 & B & Reported by \citet{viironen2009b};  visible morphology \\
CLU J195405.86$+$280740.6 & 298.52444 & 28.12795 & 64.81316 & 0.1744 & A & PN in SIMBAD reported by \citet{sanchezcontreras2008}; clear morphology \\
CLU J195707.62$+$303215.8 & 299.28175 & 30.53773 & 67.21649 & 0.85807 & B & Mid-infrared source in SIMBAD, unlabeled in HASH; visible morphology \\
CLU J195956.42$+$304823.9 & 299.98508 & 30.80663 & 67.7622 & 0.48326 & A & Reported by \citet{viironen2009a} without spectrum; SNR\,G067.8$+$0.5 nearby \\
CLU J200514.60$+$322125.1 & 301.31082 & 32.35697 & 69.67561 & 0.34923 & A & Young PN \citep{viironen2009a,corradi2010}; clear morphology \\
CLU J200850.35$+$333729.9 & 302.20978 & 33.6249 & 71.14999 & 0.39835 & B & Suspected star forming region in SIMBAD; visible morphology \\ 
CLU J201126.21$+$331606.7 & 302.85921 & 33.26853 & 71.14701 & -0.25021 & A & None PN suspected in HASH \citep{viironen2009a} without spectrum \\
CLU J201756.61$+$335718.2 & 304.48587 & 33.95506 & 72.46988 & -0.99632 & A & Extended, Possible PN reported by the amateur community
\enddata
\tablecomments{
Col. (1): Name of the object in the CLU survey.
Col. (2)--(5): Coordinates.
Col. (6): Grade of the source assigned in our visual inspection.
Col. (7): Notes on the source.
The horizontal line separates the newly identified candidates by CLU with those previously been reported by other surveys.}
\end{deluxetable*}
\end{rotatetable*}

\section{Discussion} \label{sec:discussion}
\begin{figure}[ht]
    \centering
    \includegraphics[width=\linewidth]{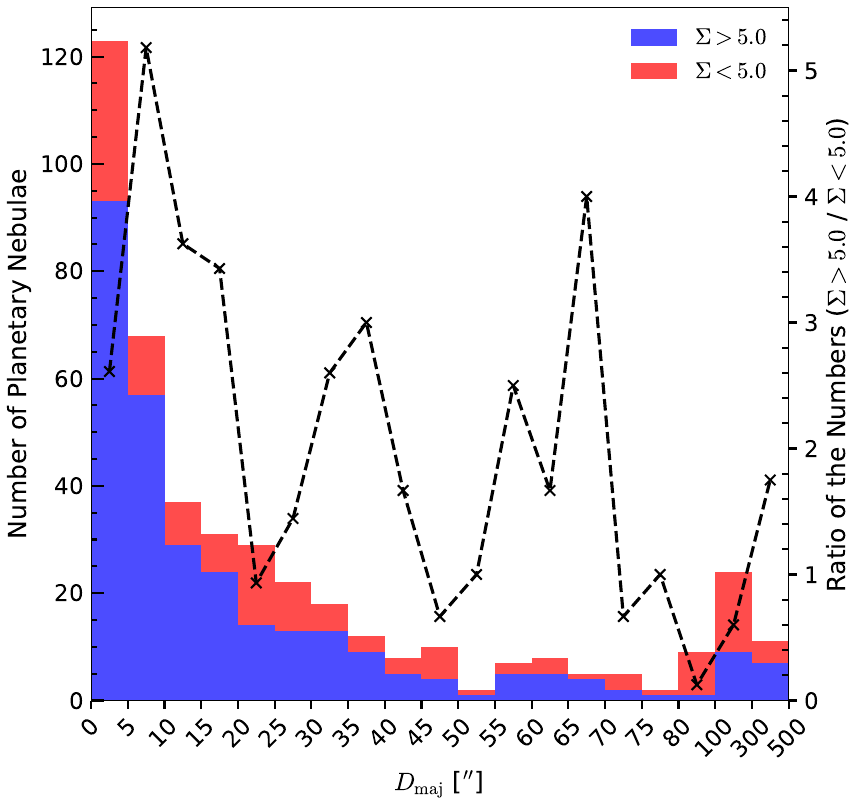}
    \caption{Stacked histogram of planetary nebulae in the validation catalog with respect to different major diameters ($D_\mathrm{maj}$). The crosses show the number ratio of candidates with $\Sigma > 5$ to $\Sigma < 5$. Sources are divided into two groups with color sigma larger or smaller than 5.}
    \label{fig:iphas_csig_d}
\end{figure}

\subsection{Caveats} \label{sec:caveats}
We note three primary caveats that are inherent in our methodology of source catalog search. First, our search for PNe focuses on H$\alpha$ excess and necessitates a minimum $m_\mathrm{H\alpha}$ threshold of 11, leading to the exclusion of those very bright nebulosities from consideration. Nonetheless, $m_\mathrm{H\alpha} = 11$ is a rather bright limit, and the majority of the brightest PNe are very likely to have already been discovered and studied. 

Second, the photometric calculations employ an aperture size of only 5 pixels ($\sim 5''$) which can lead to a possible underestimation of the color sigma because a portion of the nebular emission may not be captured. In Figure~\ref{fig:hist_filterpara}, 32\% objects in the validation catalog have $\Sigma$ smaller than 5. Interestingly, larger $D_\mathrm{maj}$ sources are in the $\Sigma < 5$ group, whereas sources with $\Sigma > 5$ are distributed more frequently with smaller $D_\mathrm{maj}$, as shown in Figure~\ref{fig:iphas_csig_d}. It is reasonable to consider using a larger aperture, but contaminating emission from nearby sources may bias the detectability of the emission line signal. Alternatively, lowering the $\Sigma$ thresholds has the risk of directly including more contaminants. If the contaminant removal techniques matures, we may adopt a lower requirement for $\Sigma$ (2.5 to 3) that still holds as a useful H$\alpha$ excess indicator. However, a bright subpart of the nebulae should be sufficient to be picked during our catalog searches, which will lead to subsequent visual inspections with zoomed-out, large FoV images for further selection. 

Finally, another issue on the large PNe is that their optical center encoded in PS1 and H$\alpha$ center from the CLU may not coincide, with $dist2stack$ easily exceeding our current cut-off value of 1.5$''$. But, previous systematic searches that relied on visual scrutiny are sensitive to large PNe, as exemplified by \citet{sabin2014} on rebinned H$\alpha - r$ difference images. Consequently, it is unlikely that there are large numbers, or any, large PNe left to be discovered. In our preliminary search, most discovered candidates are indeed compact and star-like as expected (c.f. Column~(7) of Table~\ref{tab:clupne}).

For stellar population removal (c.f. Section~\ref{sec:color_cuts}), our current cuts on the ($r - \mathrm{H}\alpha$) versus ($r - i$) diagram provides only a rough solution. \citet{viironen2009a} applied similar cuts to initially identify thousands of PN candidates from the IPHAS images. Nevertheless, subsequent spectroscopic follow-up revealed that over 80\% of these sources were still emission-line stars, which underscored the similarities in photometric properties between compact PNe and stars whereby a more reliable stellar population removal would be a difficult task. In fact, young PNe or pre-PN could be relatively red, and dust obscured sources may also be reddened, both of which add to false positives. In our preliminary search, we estimate to effectively remove $\sim 90\%$ of stars with our simple criterion. Yet, since stars roughly take up to 48\% of those $10^4$ candidate sources from the source catalog search, the remaining number of stars ($48\% \times (100\% - 90\%) = 5\%$) is still more than 15 times the number of PN candidates (0.3\%). In the future, we will improve and expand these color-color cuts to filter more stars while keeping the red PNe.

\begin{figure}[ht!]
    \centering
    \includegraphics[width=\linewidth]{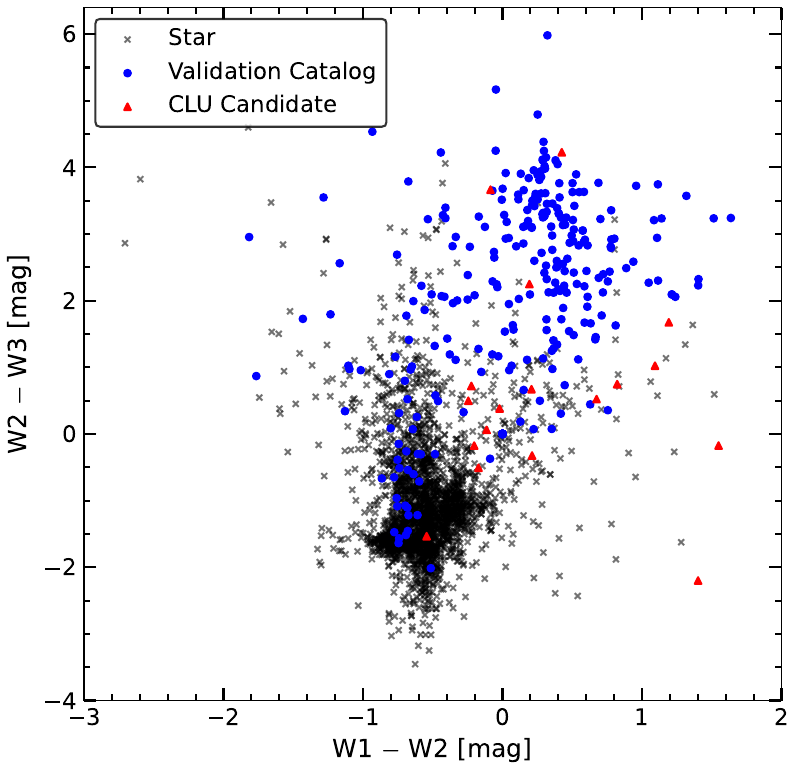}
    \caption{An example color-color diagram depicting the WISE (W2 $-$ W3) versus (W1 $-$ W2) scatter plot of labelled stars from contaminant removal, planetary nebula candidates in the validation catalog, and candidates from our preliminary search on CLU.}
    \label{fig:wise_colors}
\end{figure}

For further removal, mid-IR (MIR) and radio properties could be of help \citep[e.g.,][]{parker2012}. Since genuine PNe exhibit bremsstrahlung emission, \cite{condon1999} utilized the 1.4\,GHz NRAO VLA Sky Survey \citep[NVSS;][]{condon1998} images to identify PNe over radio-quiet MIR contaminants with the criterion of $S(25\,\mu \rm m)\gtrsim 2.5\,mJy~beam^{-1}$ by positional coincidence for a 4\% of missing rate, demonstrating that radio selection would aid on to reliable detection of PNe. Meanwhile, as adopted by HASH, the median 8\,$\mu$m MIR to radio flux ratio should be consistent with $4.7 \pm 1.1$ for all PNe \citep{cohen2011}. In addition, WISE could be used with all-sky coverage, which may help distinguishing symbiotic stars, post asymptotic giant branch stars, Wolf-Rayet shells, Mira variables and other red stars \citep[e.g.,][]{parker2012,akras2019}. Similar to the H$\alpha$-PS1 color-color cut, we briefly show the potential of WISE color cuts here. Generally, (W1 $-$ W2), (W2 $-$ W3), and (W1 $-$ W4) can all separate PN candidates from stars \citep[e.g.,][]{parker2012,akras2019}. If we adopt a very rough cut off with $\rm W1 - W2 < 0$ and $\rm W2 - W3 <1$ and $\rm W1 - W4 < 2$, 91\% stars are removed while 61\% PN candidates are kept; whereas 88\% in the validation catalog remain, since they are more extended with presumably less similarity to stars. We plot sources from the validation catalog and our preliminary search with their (W1 $-$ W2) versus (W2 $-$ W3) in Figure~\ref{fig:wise_colors} for demonstration.

Additionally, we believe a future systematic search for PNe may involve adopting a more modern and reliable detection method, such as machine learning \citep[e.g., recent attempts by][]{sun2024}, to remove all contaminants directly from candidate lists and finalize the PN candidates for spectroscopic follow-up. However, feature engineering for a machine learning classifier is complex, requiring well-established empirical expressions based on physical properties of the sources\footnote{Although deep learning methods \citep[see e.g., review by][]{lecun2015} do not require feature engineering, the interpretability of deep neural networks remains challenging.}. With prospective homogeneous sampling, the future CLU PN catalog can serve as a labeled validation dataset for emerging machine learning techniques.

Finally, we reiterate that all sources documented from the preliminary search and any catalog search of the same kind should only be considered as candidates. In general, the cuts in source catalog searches only filter out apparent emission-line excess objects that are not overly extended or bright, encompassing candidates such as some Str{\"o}mgren spheres, compact H\,{\small II} regions, Wolf-Rayet shells, SNRs, nearby emission-line galaxies, stars, image artifacts, and finally compact young PNe. Visual scrutiny is always needed to identify PN candidates, where better contaminant removal techniques will be essential to reduce the labor and boost its efficiency. Typical visual inspection can reliably discern artifacts, most Milky Way H\,{\small II} regions, and several compact stellar signals. Automatic artifact removal methods are also readily developed \citep[e.g.,][]{van-dokkum2001}, e.g., for images having sufficient quality like those with more than three images going into the stack for a single source location (Cook et al. 2024, submitted; Zhang et al. 2024, submitted). Nevertheless, spectroscopic follow-up is also indispensable for establishing a high-fidelity sample. For spectral features, H\,{\small II} regions usually exhibit [N\,{\small II}]/H$\alpha$ ratios below 0.5 to 0.7 \citep{kennicutt2000}, while non-obscured PNe tend to display higher ratios indicative of elevated gas temperatures. Bright MIR emission characterizes Wolf-Rayet shells, enabling their identification via WISE all-sky fluxes. SNRs are identifiable by their prominent sulfur and oxygen lines. Optical spectroscopy facilitates finding galaxies through their redshifted emission lines. In future work, we will propose spectroscopic follow-up for all PN candidates to confirm their status.

\begin{figure}[ht]
    \centering
    \includegraphics[width=\linewidth]{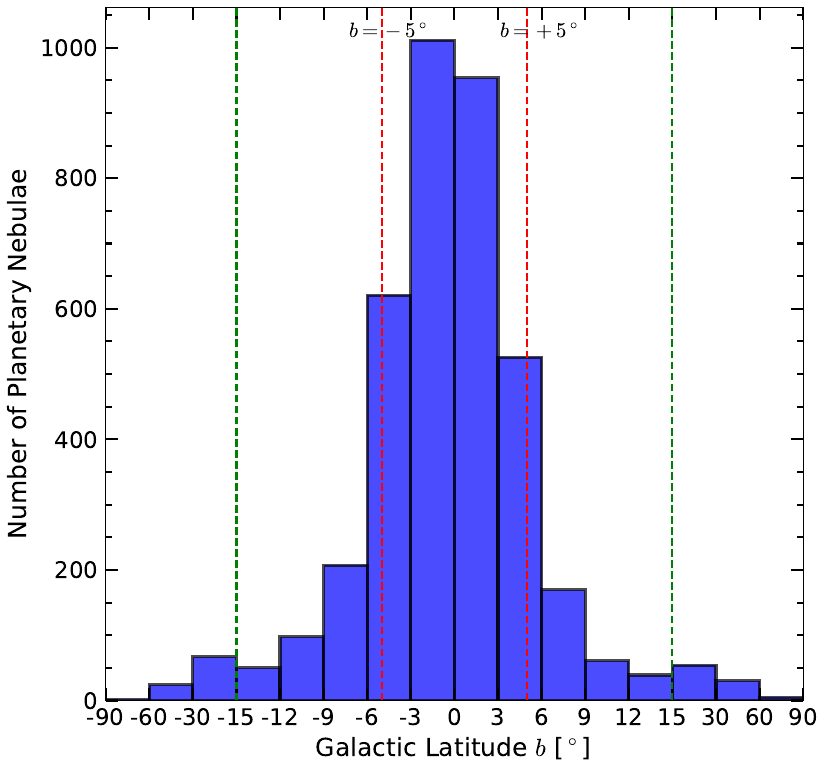}
    \caption{Histogram of the Galactic latitudes of the True, Likely, and Possible Galactic planetary nebulae in the HASH database. The bins for the latitudes are not evenly spaced.}
    \label{fig:hash_glat}
\end{figure}

\subsection{Prospects for the CLU Catalog of New Planetary Nebula} \label{sec:prospects}
The known Galactic PN population is as large as $\sim 3,900$, which is, nevertheless, below the predicted value. Depending on the formation scheme of PNe, if a more strict case (e.g., close binaries forming a common-envelope phase) is required, a theoretical value of $\sim 6,600$ \citep{demarco2005} should be expected. However, it is largely accepted that a mere 10–20\% of all PNe have binary CSPN \citep{miszalski2009}, and thus, the overall number of Galactic PNe can be up to $46,000 \pm 13,000$ based on a stellar population synthesis model \citep{moe2006}. Hence, the true number of Galactic PNe also provides a test of current demographic model for stars in our Galaxy.

Previous surveys of PNe \citep[e.g.,][]{abell1966,perek1967,acker1992,parker2005,parker2006,miszalski2008,boumis2006,gorny2006,viironen2009a,viironen2009b,sabin2014} made significant advancements in finding thousands of PNe with considerable effort at lower galactic latitudes, where the largest number of PNe are likely to be found. In addition, the amateur community has made considerable contributions to the search for PNe in other parts of the sky \citep[e.g.,][]{kronberger2012,kronberger2014,kronberger2016,acker2014,le-du2017,le-du2022}. We aim for the CLU survey to complement these previous efforts via systematic PNe searches in our survey area, where we anticipate the largest contributions to come from intermediate Galactic latitudes.

The largest Galactic plane survey in the northern hemisphere -- IPHAS -- was limited to $|b| \leq 5^\circ$, with an exclusive focus of revealing emitters near the highly obscuring Galactic plane. SHS, the largest Galactic plane search for PNe in the southern hemisphere, has a coverage of $-145^\circ < l < 30^\circ$ and mainly $|b| < 6^\circ$ with substantial coverage extending to $b = 15^\circ$. According to \citet{parker2006} and \citet{miszalski2008}, the number of SHS/MASH PNe at $|b| > 5^\circ$ declines significantly comparing to the number of PNe within $|b| = 5^\circ$ ($\sim 1000$). Yet with a detection limit down to $\sim 10^{-16} - 10^{-17} \rm \, erg~s^{-1}~cm^{-2}$, $\sim 200$ PNe are still discovered by SHS between $5^\circ < |b| < 10^\circ$, and $\sim 20$ PNe are discovered between $10^\circ < |b| < 15^\circ$. Therefore, we can expect plenty of new PNe to be discovered between $5^\circ < b < 15^\circ$, assuming that the distribution of actual PNe is close to that of the discovered population.

Since SHS was conducted on $\sim 4,000\, \rm deg^2$ of the Southern Milky Way \citep{parker2005}, and since CLU covers an area of 3,400\,deg$^2$ near Galactic Plane outside of both IPHAS and SHS, we anticipate finding around $220/4000 \times 3400 =187$ new PNe, even though SHS covers more of the Galactic bulge which may be riper for PNe with higher areal number density. Moreover, we are also likely to find a few new PNe at higher galactic latitudes. If we further refer to the the HASH database (see Figure~\ref{fig:hash_glat}), $\sim 1/3$ of known PNe are located beyond $|b| = 5^\circ$, and specifically $5\%$ of PNe are at $|b| > 15^\circ$. Thus, we may expect to find roughly $187 \times 0.05/(0.33-0.05) = 35$ high latitude PNe. Yet, since the calculation is purely based on the distribution of known PNe, the result of finding a mere 35 high latitude PNe might be underestimated if more PNe are actually distributed slightly higher in some clusters or just solely by itself.

In practice, the lower line sensitivity of CLU might not be compensated by image stacking and careful selection techniques, which could lower the expected numbers of CLU PNe. However, despite of a brighter detection limit of the CLU ($1.6 \times 10^{-14} \rm \, erg~s^{-1}~cm^{-2}$, Cook et al. 2024; submitted), we are already finding new candidates in the same fields as those of deeper searches. Meanwhile, the CLU survey will also face a lower contamination level at higher latitudes where the easily removable nearby galaxies could become the major false positive. After the entire catalog search with automatic and visual contaminant removal procedures, we anticipate to find approximately $187+35 \approx 220$ new PNe only through scaling by areas. Other practical factors may affect the final number of CLU PNe, which we leave as a topic for future work.

\section{Conclusion} \label{sec:conclusion}
Planetary nebulae play a pivotal role in deciphering the intricacies of stellar evolution, Galactic chemical enrichment, and broader cosmological phenomena. Leveraging data from the CLU narrowband H$\alpha$ emission-line galaxy survey, our project aims to systematically detect and catalog PNe within the CLU survey extending above a declination of $-20^\circ$.

Originally targeting H$\alpha$ galaxies, the CLU survey has proven its efficacy in PN searches with a 98\% recovery rate of 441 PNe within a validation catalog constructed from IPHAS/HASH data. The quality and utility of CLU were shown by the fact that 90\% of these sources exhibit readily discernible morphology in CLU-H$\alpha$ images. Notably, our recovery rates remained consistently high across sources of all HASH PN classes, with 95\% of True, 71\% of Likely, and 81\% of Possible sources being readily identified.

We conducted a preliminary search for PN candidates, employing a source catalog search primarily based on color sigma ($\Sigma > 5$) supplemented by visual inspection for contaminant removal. Within the region of $|b| < 2^\circ$ and $25^\circ < l < 75^\circ$, we discovered 31 PN candidates where 12 are new sources that have not appeared in any previous studies. Spectroscopic follow-up has been conducted for six candidates so far, revealing that four of them are good candidates and two of them are likely contaminants.

Spanning intermediate to high Galactic latitudes, the CLU survey presents a unique opportunity for uncovering PNe in previously unexplored regions of the sky. Looking ahead, our future work will be geared towards refining catalog search criteria and conducting spectroscopic follow-up to validate and confirm additional candidates.

\section*{Acknowledgments}

We thank the anonymous referee for helpful suggestions. R.D. thanks Luis C. Ho and Yuanze Ding for useful suggestions in the early stage of the project. S.B. thanks Howard E. Bond for useful discussion on the DBSP spectra of the planetary nebula candidates. 

We make use of the Intermediate Palomar Transient Factory project which is a scientific collaboration among the California Institute of Technology, Los Alamos National Laboratory, the University of Wisconsin, Milwaukee, the Oskar Klein Center, the Weizmann Institute of Science, the TANGO Program of the University System of Taiwan, and the Kavli Institute for the Physics and Mathematics of the Universe. 

This research has made use of the HASH PN database at hashpn.space. The HASH database is maintained by the Laboratory for Space Research at the University of Hong Kong. 

This research has made use of the SIMBAD database, operated at CDS, Strasbourg, France. This research has made use of the NASA/IPAC Extra-galactic Database (NED), which is operated by the Jet Propulsion Laboratory, California Institute of Technology, under contract with the National Aeronautics and Space Administration. 

The Pan-STARRS1 Surveys (PS1) and the PS1 public science archive have been made possible through contributions by the Institute for Astronomy, the University of Hawaii, the Pan-STARRS Project Office, the Max-Planck Society and its participating institutes, the Max Planck Institute for Astronomy, Heidelberg and the Max Planck Institute for Extra-terrestrial Physics, Garching, Johns Hopkins University, Durham University, the University of Edinburgh, the Queen's University Belfast, the Harvard-Smithsonian Center for Astrophysics, the Las Cumbres Observatory Global Telescope Network Inc., the National Central University of Taiwan, the Space Telescope Science Institute, the National Aeronautics and Space Administration under grant no. NNX08AR22G issued through the Planetary Science Division of the NASA Science Mission Directorate, the National Science Foundation Grant no. AST-1238877, the University of Maryland, Eotvos Lorand University (ELTE), the Los Alamos National Laboratory, and the Gordon and Betty Moore Foundation. 

Funding for SDSS-III has been provided by the Alfred P. Sloan Foundation, the Participating Institutions, the National Science Foundation, and the U.S. Department of Energy Office of Science. The SDSS-III website is \url{http://www.sdss3.org/}. SDSS-III is managed by the Astrophysical Research Consortium for the Participating Institutions of the SDSS-III Collaboration, including the University of Arizona, the Brazilian Participation Group, Brookhaven National Labora- tory, Carnegie Mellon University, the University of Florida, the French Participation Group, the German Participation Group, Harvard University, the Instituto de Astrofisica de Canarias, the Michigan State/Notre Dame/JINA Participation Group, Johns Hopkins University, Lawrence Berkeley National Laboratory, Max Planck Institute for Astrophysics, Max Planck Institute for Extraterrestrial Physics, New Mexico State University, New York University, Ohio State University, Pennsylvania State University, the University of Portsmouth, Princeton University, the Spanish Participation Group, the University of Tokyo, the University of Utah, Vanderbilt University, the University of Virginia, the University of Washington, and Yale University.

%

\vspace{5mm}
\facilities{48-inch Oschin, 2.5\,m INT, 200-inch Hale}


\software{\textsc{AstroPy} \citep{astropy2013,astropy2018,astropy2022},
        \textsc{Matplotlib} \citep{hunter2007},
        \textsc{Numpy} \citep{van-der-walt2011,harris2020},
        \textsc{SciencePlots} \citep{garrett2021},
        \textsc{TOPCAT} \citep{taylor2005}
          }




\appendix

\section{Recovery Results of the IPHAS Catalog} \label{app:iphas_result}
In Section~\ref{sec:rediscov}, we discussed the visual recovery process applied to a validation catalog constructed from IPHAS/HASH data. Table~\ref{tab:recover_iphas} presents our grading in Column~(9) and observational notes for each candidate in Column~(10), as well as detailed information for each source collected from previous surveys.

\section{DBSP Spectra of Six PN Candidates}\label{app:dbsp_spectra}
We provide the DBSP optical spectroscopic observations of the six PN candidates. In all the Figures, the upper and lower left panels show the CLU H$\alpha$1 ``On'' and H$\alpha$2 ``Off'' filter images, respectively. The upper middle and right panels show the two-dimensional spectra from DBSP in the blue and red channels respectively. The lower right panel shows the one-dimensional spectrum extracted from the marked trace. The unreliable (elevated noise, prominent cosmic rays etc.) regions of the spectrum have been shaded.

\begin{longrotatetable}
\begin{deluxetable*}{ccccccccccccc}
\tablenum{A1}
\tablecaption{Results of Recovering the IPHAS PNe in the CLU Imaging Data \label{tab:recover_iphas}}
\setcounter{table}{2}
\setlength{\tabcolsep}{3pt}
\movetabledown=12mm
\tablehead{
\colhead{Name} &
\colhead{R. A. (J2000)} &
\colhead{Decl. (J2000)} &
\colhead{$D_\mathrm{maj}$} &
\colhead{$D_\mathrm{min}$} &
\colhead{M. Mor.} &
\colhead{S. Mor.} &
\colhead{Class} &
\colhead{Grade} &
\colhead{Notes} &
\colhead{$l$} & 
\colhead{$b$} & 
\colhead{Ref.}
\\ 
&
\colhead{($^\mathrm{h} \, ^\mathrm{m} \, ^\mathrm{s}$)} &
\colhead{($^\circ \, ' \, ''$)} &
\colhead{($''$)} &
\colhead{($''$)} &
&
&
&
& 
& 
\colhead{($^\circ$)} &
\colhead{($^\circ$)} 
}
\decimalcolnumbers
\startdata
PM 1-246                  & 18 36 02.4 & -00 02 27.1 & 16      & 16      & E         &          & L    & C       & Small; Faint                       & 31.13806  & 3.40687  & 1         \\
PM 1-262                  & 18 53 02.0 & +06 44 15.3 &         &         & S         &          & T    & A       &                                    & 39.11395  & 2.71376  & 1         \\
PM 2-40                   & 19 17 50.6 & +08 15 08.3 &         &         & R         &          & L    & A       &                                    & 43.28616  & -2.0518  & 1         \\
PM 1-302                  & 19 27 51.4 & +14 01 27.8 & 10      &         & R         & ars      & T    & A       &                                    & 49.5371   & -1.49728 & 1         \\
PM 1-305                  & 19 31 41.3 & +22 43 38.8 & 15      &         & E         & as       & T    & A       &                                    & 57.61212  & 1.87629  & 1         \\
PM 1-319                  & 20 10 37.7 & +31 37 56.6 &         &         & S         &          & L    & A       &                                    & 69.68419  & -1.00291 & 1         \\
PM 1-332                  & 21 34 23.2 & +52 37 27.6 & 6.5     & 5.7     & E         &          & T    & A       &                                    & 95.54608  & 0.56095  & 1         \\
BV 5-1                    & 00 19 58.7 & +62 59 01.7 & 42      & 10      & B         & ms       & T    & A       &                                    & 119.368   & 0.32993  & 2         \\
\enddata
\tablecomments{
Col. (1): Name of the object.
Col. (2)--(3): Coordinates. 
Col. (4)--(5): Diameter of the major and minor axis of the nebula.
Col. (6)--(7): Main and Sub morphological type.
Col. (8): Planetary nebula class with ``T'' meaning ``True'', ``L'' meaning ``Likely'', and ``P'' meaning ``Possible''.
Col. (9): Grade of the source assigned in our visual recovery.
Col. (10): Notes on the source from our visual recovery.
Col. (11)--(12): Galactic longitude and latitude.
Col. (13): Reference catalog of the source. 
}
\tablerefs{1--\citet{preite-martinez1988}; 2--\citet{acker1992}; 3--\citep{kohoutek2001}; 4--\citep{kerber2003}; 5--\citep{kronberger2006}; 6--\citep{mapaso2006}; 7--\citep{parker2006}; 8--\citep{miszalski2008}; 9--\citep{wesson2008}; 10--\citep{miranda2009}; 11--\citep{urquhart2009}; 12--\citep{viironen2009a}; 13--\citep{viironen2009b}; 14--\citep{jacoby2010}; 15--\citep{sabin2011}; 16--\citep{viironen2011}; 17--\citep{acker2012}; 18--\citep{kronberger2014}; 19--\citep{sabin2014}; 20--\citep{kronberger2016}; 21--\citep{le-du2017}; 22--\citep{nunez2017}; 23--\citep{fragkou2018}; 24--uncataloged objects discovered with the Digital Sky Hunter \citep[DSH; see proceedings associated with e.g., ][]{kronberger2006,jacoby2010}; 25--uncataloged objects discovered by the amateur community; 26--uncataloged objects discovered as candidates in IPHAS.}
\end{deluxetable*}
\end{longrotatetable}

\begin{figure*}[t!]
\figurenum{B1}
    \centering
    \includegraphics[width=\linewidth]{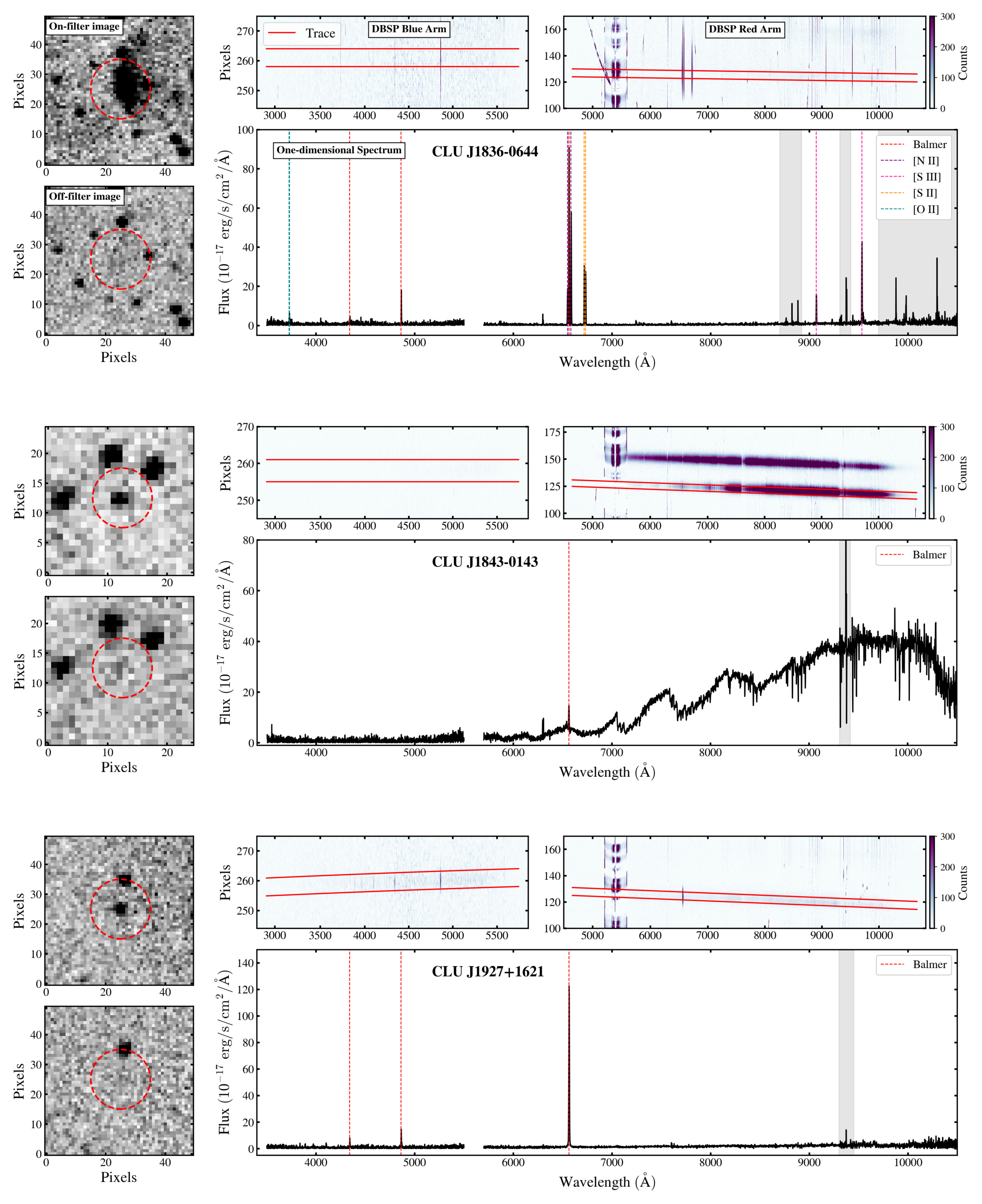}
    \caption{DBSP follow-up spectra for (from top to bottom) CLU\,J183648.44$-$064442.8, CLU\,J184324.79$-$014319.2, and CLU\,J192710.49$+$162128.6. In each figure, \textbf{Left panel:} shows the CLU images (top: ON/H$\alpha$ filter, bottom: OFF filter), \textbf{Two upper-right panels:} two-dimensional DBSP spectra for the blue and red channels respectively, with the trace marked, and \textbf{Lower-right panel:} the one-dimensional spectra extracted from the marked trace. The identified prominent are marked and the contaminated portions of the spectra are shaded. We have omitted the extreme ends of the blue and red arms ($\simeq$$5400-5800$~\AA) due to significantly elevated noise in this break region.}
    \label{fig:combined1}
\end{figure*}

\begin{figure*}[t!]
\figurenum{B1 continued}
    \centering
    \includegraphics[width=\linewidth]{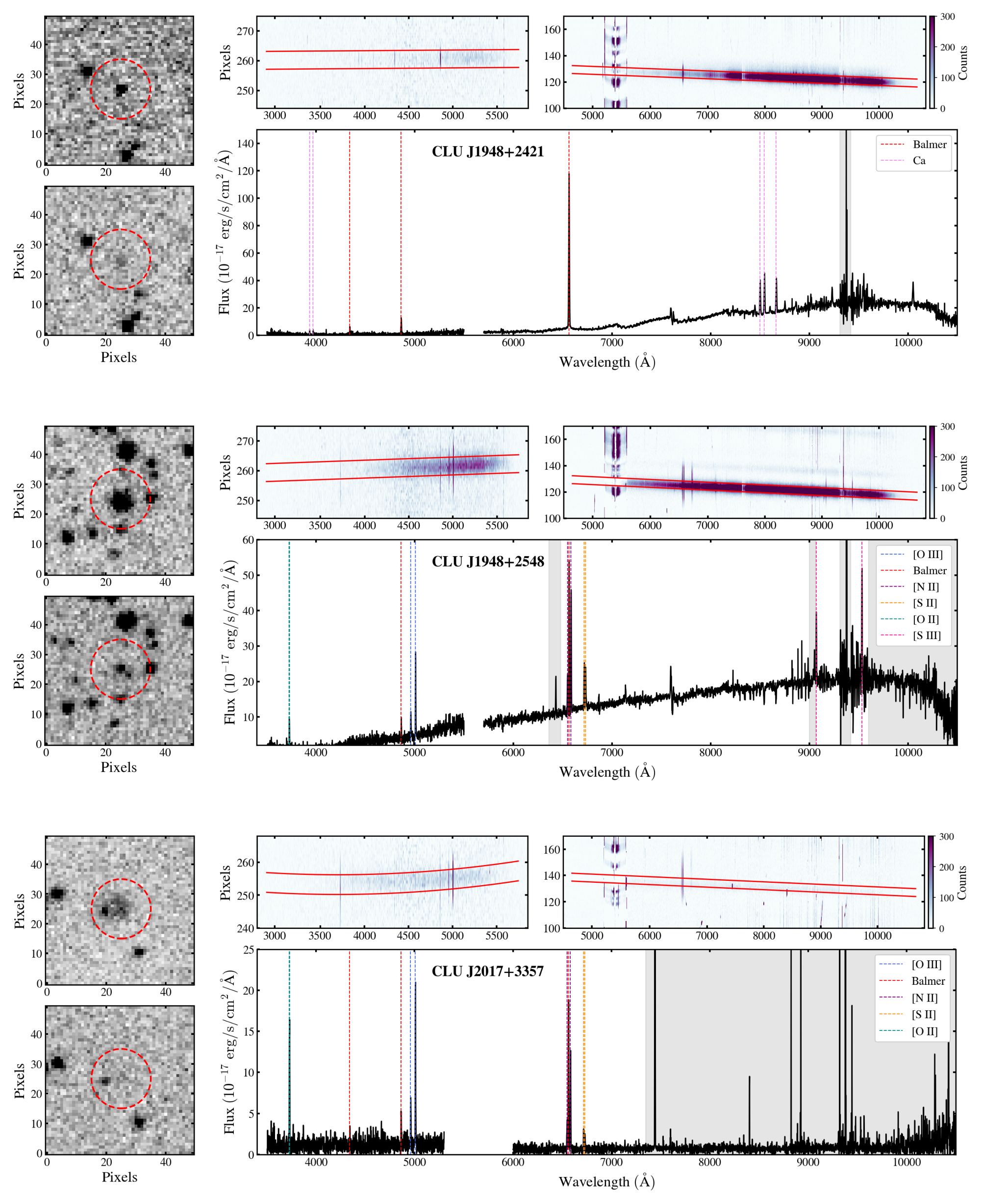}
    \caption{DBSP follow-up spectra for (from top to bottom) CLU\,J194817.60$+$242116.7, CLU\,J194804.17$+$254848.6, and CLU\,J201756.61$+$335718.2. Same format as Figure \ref{fig:combined1}.}
    \label{fig:combined2}
\end{figure*}

\bibliography{pne}{}
\bibliographystyle{aasjournal}
\end{document}